\documentclass[a4paper, 12pt]{article}
\usepackage{eurosym}
\usepackage{amssymb}
\usepackage{amsmath}
\usepackage{graphicx}

\input{tcilatex}
\begin{document}

\title{Effects of the parallel acceleration on \\
heavy impurity transport in turbulent tokamak plasmas}
\author{Madalina Vlad, Dragos Iustin Palade, Florin Spineanu \\
National Institute of Laser, Plasma and Radiation Physics, Bucharest, Romania%
}
\maketitle

\abstract{
A process specific to the dynamics of the heavy impurities in turbulent tokamak plasmas is found and analysed. We show that the parallel stochastic acceleration is strongly coupled to the perpendicular transport and generates a radial pinch velocity. The interaction is produced with the hidden drifts, a quasi-coherent component of the motion that consists of a pair of average radial velocities in opposite directions. The parallel acceleration breaks this symmetry and yields a radial average velocity that can be in the inward or outward direction. The transport of the tungsten ions in three-dimensional turbulence is analysed in the frame of a test particle model using numerical simulations. The results show that the acceleration induced pinch can be important for W impurity transport in present days tokamaks and in ITER.
}

\bigskip

\section{Introduction}

Tungsten (W) will be used for plasma facing components in ITER, because this
material fulfils essential requirements such as low erosion rate, low
tritium retention and good thermal properties \cite{W}. The effects of the W
wall on tokamak plasma performances have been intensively studied during the
last decades in several tokamak devices (ASDEX Upgrade \cite{Neu}-\cite{Bock}%
, JET \cite{Putterich}-\cite{Valisa}, WEST \cite{Bucalossi}, T-10 \cite%
{Krupin17}, JT-60U \cite{Nakano}). The main drawback is the large radiation
emission of these high charge ions that can strongly affect the energy
balance if they accumulate in the core plasma in concentrations higher that $%
10^{-5}-10^{-4}$. It is, therefore, vital to acquire a good understanding of
impurity transport and to develop methods for controlling the concentration
of W in tokamak core plasmas. A large number of experimental \cite{Sertoli}-%
\cite{Kochl18}, theoretical \cite{Angioni}-\cite{Breton18} and numerical 
\cite{Casson15}-\cite{Yamoto} studies have provided important results on
these complex processes, but the domain is still open as shown by very
recent papers \cite{Casson20}-\cite{Ferrari19}

The dynamics of the W ions includes both neoclassical and turbulent effects.
The large mass and charge determine strong inertial and electrostatic
forces, with the result of specific phenomena that are not observed at light
ions. Strong poloidal asymmetries, significant increase of the neoclassical
transport and accumulation on the low field side of the plasma or even
around the magnetic axis characterize W ions. Radial convections with
neoclassical or turbulence origins can determine W accumulation or decay.

\bigskip

The present paper deals with the turbulent transport of the heavy
impurities. We analyze the effects of the parallel acceleration $a_{z}$ on
the diffusion coefficients and on the radial pinch velocity. The
acceleration scales as $a_{z}\sim Z/A,$ where $Z$ is the ionization rate\
and $A$ is the mass number of the ions. Thus, it is smaller for the W
impurities than for plasma ions. The factor $Z/A$ varies in the interval $%
\left( 0.05,\ 0.33\right) $ for the $W$ impurities, while it has the value $%
0.5$ for the deuterium.

Surprisingly, we have found that the effect of $a_{z}$ can be significant
for $W$ impurities, while it is negligible for plasma ions.

The main effect of the parallel acceleration consists of the generation of a
radial pinch velocity $V_{x}$. The aim of the paper is to understand and to
characterize this new mechanism of generating radial pinch.

The parallel acceleration is expected to influence impurity transport
through the modification of the parallel decorrelation time. We show that,
beside this direct effect, a much stronger coupling of the parallel
accelerated motion to the radial transport appears. It consists of the
perturbation of hidden drifts (HDs). The HDs are a pair of opposite
velocities in the radial direction that appear in the presence of a poloidal
average velocity \cite{VS18}. This quasi-coherent motion has zero average
and does not determine a convective velocity in the case of the ExB drift.
The stochastic parallel acceleration perturbs the equilibrium of the HDs
leading to a radial pinch. We show that this pinch mechanism can be relevant
for W impurity accumulation in present days and in ITER plasmas.

The analysis is performed in the frame of a test particle stochastic model,
which is shown to be the minimal model that yields this process. The model
is presented in Section 2.

We use two theoretical methods for determining the pinch velocity and the
transport coefficients, the direct numerical simulations (DNS) \cite%
{Palade20} and the decorrelation trajectory method (DTM) \cite{Vlad98}. A
three-dimensional DNS code for ion trajectories and for the calculation of
the statistical Lagragian quantities was developed. It is described in
Section 3.1. The DTM is a semi-analytical approach that provides approximate
evaluations of the transport characteristics. It is presented in Section
3.2. The DTM is used for identifying and understanding qualitatively the new
pinch mechanism, while the quantitative properties of the radial velocity $%
V_{x}$ are determined using the much more accurate results provided by DNS.\
\ 

The effects of the parallel acceleration on the heavy impurity transport are
identified in Section 4 by comparing typical results of the transport model
with those obtained for deuterium ions and for W ions in two-dimensional
potentials.

The physical processes that determine the generation of the radial pinch are
discussed in Section 5. We use the DTM, which has the capability to provide
physical pictures of complex nonlinear transport processes \cite{VS13}-\cite%
{VS16}. We show that the acceleration can have a strong influence on the HDs
that essentially consists of the attenuation of one of the HDs, which
compensates only partially the other HD yielding an average velocity. The
physical image of the pinch generation mechanism is validated using DNS. A
short discussion on the accuracy of the DTM is also presented in this
Section.

The properties of the pinch velocity $V_{x}$ and its dependence on the main
parameters of the model are determined in Section 6. They are obtained using
the more accurate results of the DNS. The study is focused on the scaling of 
$V_{x}$ with the main parameters of the model. The results are analyzed and
physical explanations are derived.

The relevance of the pinch generated by the parallel acceleration for the W
ion transport in the existing plasmas (ASDEX Upgrade and JET) and in ITER is
discussed in Section 7. A summary of the results and the conclusions of this
study are also included in this section.

\section{The transport model}

We study impurity transport in the slab approximation, at the low field side
of the plasma. The magnetic field is constant along $\mathbf{e}_{z}$ axis
and $\mathbf{x=}\left( x,y\right) $ is in the perpendicular plane, with $x$
the radial and $y$ the poloidal coordinate. The equations for the impurity
ion trajectories are%
\begin{equation}
\frac{d\mathbf{x}}{dt}=-\frac{\mathbf{\nabla }\phi \times \mathbf{e}_{z}}{B}+%
\mathbf{V}_{d},  \label{eqm}
\end{equation}%
\begin{equation}
\frac{dz}{dt}=v_{z},~\ \frac{dv_{z}}{dt}=-\frac{q}{m}\partial _{z}\phi ,
\label{eqmz}
\end{equation}%
where the first term in Eq. (\ref{eqm}) is the stochastic drift determined
by the electric field of the turbulence $-\mathbf{\nabla }\phi (\mathbf{x,}%
z,t)$ ($\phi (\mathbf{x,}z,t)$ is the stochastic potential, $\mathbf{\nabla }
$ is the gradient in the perpendicular plane) and the second term $\mathbf{V}%
_{d}=V_{d}\mathbf{e}_{y}$ is a poloidal average velocity that can be
produced by the magnetic drifts or plasma rotation. The parallel motion (\ref%
{eqmz}) includes the variation of the velocity determined by the stochastic
acceleration $a_{z}=-q/m\ \partial _{z}\phi .$ \ 

The potential $\phi (\mathbf{x,}z,t)$ is modelled as a Gaussian random field
with the Eulerian correlation (EC)%
\begin{equation}
E(\mathbf{x,}z,t)\equiv \left\langle \phi (\mathbf{0,}0,0)\ \phi (\mathbf{x,}%
z,t)\right\rangle  \label{ECdef}
\end{equation}%
corresponding to drift type turbulence \cite{VS13}-\cite{VS16}

\begin{equation}
E(\mathbf{x,}z,t)=A_{\phi }^{2}\partial _{y}\left[ \exp \left( -\frac{x^{2}}{%
2\lambda _{x}^{2}}-\frac{y^{2}}{2\lambda _{y}^{2}}-\frac{z^{2}}{2\lambda
_{z}^{2}}\right) \frac{\sin \left( k_{0}y\right) }{k_{0}}\right] T(t),
\label{EC}
\end{equation}%
where $A_{\phi }$\ is the amplitude of the potential fluctuations, $\lambda
_{x},$ $\lambda _{y},~\lambda _{z}$\ are the correlation lengths along the
radial, poloidal and parallel directions, and $k_{0}$\ is the dominant wave
number. The function $T(t)$\ is the time correlation of the potential that
is a decaying function of time with $\tau _{d}$ the decorrelation time 
\begin{equation}
T(t)=\exp \left( -\frac{t^{2}}{4\tau _{d}^{2}}\right) .  \label{exp-dec}
\end{equation}%
\ \ 

Dimensionless quantities are used, with the units: $\rho _{i}=v_{thi}/\Omega
_{i},$\ the Larmor radius of the protons (for the perpendicular distances,
for the correlation lengths $\lambda _{x},$ $\lambda _{y}$ and for $%
1/k_{0}), $ $a,$ the small radius of the plasma (for the parallel distances
and for the correlation length $\lambda _{z}),$ $\tau _{0}=a/v_{thi}$\ (for
time and for $\tau _{d}$), $A_{\phi }$ (for the potential $\phi $), $V_{\ast
}=\rho _{i}v_{thi}/a$ (for the perpendicular velocities and $V_{d})$ and $%
v_{thW}=v_{thi}/\sqrt{A}$ (for the parallel velocity of the W ions). $%
v_{thi}=\sqrt{T_{i}/m_{p}}$ is the thermal velocity of protons with
temperature $T_{i}$ and mass $m_{p}$ and $\Omega _{i}=eB/m_{p}$ is the
cyclotron frequency of the protons.\ The notations are not changed for the
dimensionless quantities, and Eqs. (\ref{eqm})-(\ref{eqmz}) for ions with
mass number $A$ and ionization rate $Z$ are%
\begin{eqnarray}
\frac{dx}{dt} &=&-K_{\ast }\partial _{y}\phi (\mathbf{x,}z,t),\ \frac{dy}{dt}%
=K_{\ast }\partial _{x}\phi (\mathbf{x,}z,t)+V_{p},  \label{eqm1} \\
\frac{dz}{dt} &=&\frac{1}{\sqrt{A}}v_{z},\ \ \frac{dv_{z}}{dt}%
=-P_{a}\partial _{z}\phi .  \label{eqz1}
\end{eqnarray}%
The main characteristics of the model appear in three dimensionless
parameters evidenced in the dimensionless equation. The parameter $K_{\ast }$
is the dimensionless measure of turbulence amplitude%
\begin{equation}
K_{\ast }=\Phi \frac{a}{\rho _{i}},\ \ \Phi =\frac{eA_{\phi }}{T_{i}}.
\label{Kstar}
\end{equation}%
The parameter of the poloidal velocity $V_{p}$ is%
\begin{equation}
V_{p}\equiv \frac{V_{d}}{V_{\ast }}=\frac{V_{d}}{v_{thi}}\frac{a}{\rho _{i}}
\label{Vd}
\end{equation}%
The parameter of the parallel acceleration $a_{z}$ is 
\begin{equation}
P_{a}\equiv \Phi \frac{Z}{\sqrt{A}}.  \label{Pac}
\end{equation}%
We note that the first two parameters that describe the perpendicular motion
depend on plasma size factor $\rho _{\ast }=\rho _{i}/a.$

The energy of the ions normalized with the temperature $T_{i}$ is%
\begin{equation}
W=\frac{1}{2}v_{z}^{2}+Z\Phi \phi .  \label{W}
\end{equation}%
It is the invariant of the motion in three-dimensional static potentials $%
\phi (\mathbf{x,}z).$\ This constraint influences the transport for $\tau
_{d}\rightarrow \infty ,$\ and its effects persist in the case of potentials
with slow time variation (large $\tau _{d}).$\ We note that the energy is
dominated by the potential energy for the W ions with large $Z,$\ even at
small turbulence amplitudes ($\Phi \sim 10^{-2}).$

\section{Theoretical methods}

The model is analysed using direct numerical simulations (DNS) \cite%
{Palade20} and the decorrelation trajectory method (DTM) \cite{Vlad98}.

\subsection{DNS numerical methods and code}

The numerical methods used in the DNS code are described and analyzed in 
\cite{Palade20}. A series of fast numerical generators of Gaussian random
fields with given EC are proposed. In the present work, we have implemented
the so called FRD representation%
\begin{equation}
\phi (\mathbf{X})=\sum\limits_{i=1}^{N_{c}}\sqrt{S(\mathbf{K}_{i})}\sin
\left( \mathbf{K}_{i}\mathbf{X}+\frac{\pi }{4}\zeta _{i}\right) ,
\label{FRD}
\end{equation}%
where $\mathbf{X}\equiv (\mathbf{x,}z,t)$ is the four-dimensional
space-time, $\mathbf{K}_{i}\equiv (\mathbf{k}_{\perp }\mathbf{,}k_{z},\omega
)$\ are the $N_{c}$ discrete values of the corresponding wave numbers and
frequency and $S(\mathbf{K})$ is the spectrum of the stochastic potential
(the Fourier transform of the EC (\ref{EC})). This representation is
different of the usual discrete Fourier decomposition by the set of the
values of $\mathbf{K}_{i}$\ that are not the fixed points of a
four-dimensional mesh, but random values with uniform distribution. Also,
the random phases do not have continuous distributions, but discrete values $%
\pm 1$ (with equal probabilities). Each set of the $N_{c}$ random values $%
\zeta _{i}$\ determines a realization of the potential, which constitutes
the statistical ensemble (with a number $M$ of elements).

We have shown \cite{Palade20} that the representation (\ref{FRD}) provides
fast convergence of the Eulerian statistics of the generated fields, as well
as of the Lagrangian statistics of trajectories. In particular, it was
proven that a convergence level with a few percents error can be achieved
with $N_{c}\sim 10^{d}$ and $M\sim 10^{4},$ where $d=4$ for time dependent
potentials and $d=3$ for $\tau _{d}\rightarrow \infty .$ Also, it is worth
mentioning that such representations are able to reproduce with high
accuracy the conservation laws of motion as well as certain Lagrangian
statistical invariants.

The properties of the representation (\ref{FRD}) enables to use commonly in
the present simulations $N_{c}\sim 1-5\times 10^{3}$ partial waves. The
dimension of the statistical ensemble is usually set to $M\sim 10^{5}$
realizations which gives negligible statistical fluctuations. The numerical
integration scheme used is a forth order Runge-Kutta method which preserves
well the energy with a minimal numerical effort. Depending on the
integration time and on the type of turbulence (frozen, or not), the usual
CPU times on personal computer are $t_{CPU}\sim 2-20$ hours per run.

\subsection{DTM semi-analytical method}

The DTM is a semi-analytical approach, which is able to describe both the
random and the quasi-coherent components of the trajectories. The latter are
determined by the finite correlation lengths of the stochastic potential and
depend on the structure of the correlated zone that is described by the
shape of the EC.

The statistical ensemble of stochastic potentials is divided in subensembles 
$S$ with given values of the potential and of its derivatives at the origin
of the trajectories, $\mathbf{x=0,}$ $z=0,$ $t=0$ 
\begin{equation}
\phi (\mathbf{0,}0,0)=\phi ^{0},\ \partial _{i}\phi (\mathbf{0,}0,0)=\phi
_{i}^{0},~\   \label{S}
\end{equation}%
where $i=x,y,z$. The potential and its derivatives, restricted at the
realizations contained in a subensemble, are Gaussian fields with
space-dependent averages%
\begin{equation}
\left\langle \phi (\mathbf{x,}z,t)\right\rangle _{S}\equiv \Phi ^{S}(\mathbf{%
x,}z\mathbf{,}t)=\phi ^{0}\frac{E(\mathbf{x,}z\mathbf{,}t)}{E(\mathbf{0,}0%
\mathbf{,}0)}-\sum_{i}\phi _{i}^{0}\frac{E_{i}(\mathbf{x,}z\mathbf{,}t)}{%
E_{ii}(\mathbf{0,}0\mathbf{,}0)},  \label{FiS}
\end{equation}%
\begin{equation}
\left\langle \partial _{i}\phi (\mathbf{x,}z,t)\right\rangle _{S}=\partial
_{i}\Phi ^{S}(\mathbf{x,}z\mathbf{,}t),  \label{derFiS}
\end{equation}%
where $E_{ij}$ are derivatives of the EC, $E_{i}=\partial _{i}E,$ $%
E_{ii}=\partial _{i}^{2}E$. The amplitudes of fluctuations in a subensemble
vanishes in $\mathbf{x=0,}$ $z=0,$ and they reach the level corresponding to
the whole set of realizations only at large distances compared to the
correlation lengths.

Particle trajectories are studied separately in each subensemble $S$. The
average potential (\ref{FiS}), determined by the EC, yields an average
trajectory in each subensemble. It is obtained by averaging Eqs. (\ref{eqm1}%
), (\ref{eqz1}) over the realizations that belong to $S$. Neglecting the
fluctuations of the potential in $S$ (see \cite{Vlad98}, \cite{VS17} for the
discussion of this approximation), one obtains a system of subensemble
average equations (S-eq). It has the same structure as Eqs. (\ref{eqm1}), (%
\ref{eqz1}), but with the stochastic potential $\phi (\mathbf{x,}z,t)$
replaced by the subensemble average potential $\Phi ^{S}(\mathbf{x,}z\mathbf{%
,}t).$ The solution of the (S-eq), $\mathbf{X}(t;\phi ^{0},\phi _{i}^{0}),$ $%
Z(t;\phi ^{0},\phi _{i}^{0}),$ is a smooth, simple trajectory, which is
named decorrelation trajectory (DT)\ because it represents the average
evolution of the particles through the correlated zone of the potential. An
important feature of the DTs is that they obey any conservation law which
characterize the real trajectories. In our particular case, the energy 
\begin{equation}
\left\langle W(t)\right\rangle _{S}=\frac{1}{2}\left\langle
v_{z}^{2}(t)\right\rangle _{S}+Z\Phi \Phi ^{S}(\mathbf{X}(t;\phi ^{0},\phi
_{i}^{0})\mathbf{,}Z\mathbf{(}t;\phi ^{0},\phi _{i}^{0}))  \label{WL}
\end{equation}%
is conserved along the DTs for static potentials.

The statistical characteristics of the stochastic trajectories are obtained
as weighted averages along the DTs by summing the contributions of all
subensembles. In particular, the time dependent diffusion coefficient and
the average radial displacement are%
\begin{eqnarray}
D_{x}(t) &=&\int d\phi ^{0}d\phi _{x}^{0}d\phi _{y}^{0}d\phi _{z}^{0}\
P(\phi ^{0},\phi _{i}^{0})v_{x}^{0}X(t;\phi ^{0},\phi _{i}^{0}),  \label{Dx}
\\
X(t) &\equiv &\left\langle x(t)\right\rangle =\int d\phi ^{0}d\phi
_{x}^{0}d\phi _{y}^{0}d\phi _{z}^{0}\ P(\phi ^{0},\phi _{i}^{0})X(t;\phi
^{0},\phi _{i}^{0}),  \label{xme}
\end{eqnarray}%
where $P(\phi ^{0},\phi _{i}^{0})$ is the (Gaussian) probability of the
initial conditions (\ref{S}) and $v_{x}^{0}=-K_{\ast }\phi _{y}^{0}$ is the
initial radial velocity.

The average displacement generates an average velocity for stochastic
processes that have finite decorrelation times. The parallel motion provides
an intrinsic decorrelation process that yields from the $z-$dependence of
the EC

\begin{equation}
\tau _{z}^{eff}(t)=\int\limits_{0}^{t}d\tau ~\left\langle \exp \left( -\frac{%
z^{2}\left( \tau \right) }{2\lambda _{z}^{2}}\right) \right\rangle .
\label{tauef}
\end{equation}%
This is an increasing function of time that saturates at a finite value $%
\tau _{z}^{\infty }.$ The radial pinch velocity is evaluated as%
\begin{equation}
V_{x}(t)=\frac{\left\langle x\left( \tau _{z}^{eff}(t)\right) \right\rangle 
}{\tau _{z}^{eff}(t)},\ \ V_{x}^{\infty }=\frac{\left\langle x\left( \tau
_{z}^{\infty }\right) \right\rangle }{\tau _{z}^{\infty }}.  \label{Va}
\end{equation}%
\ We note that, in time dependent potentials, $\tau _{z}^{eff}(t)$\ combines
with the time dependence of the EC $T(t)$\ in a modified function $\tau
^{eff}(t)$ with a modified asymptotic value $\tau ^{\infty }.$ \ 

The time dependent functions $D_{x}(t)$ (\ref{Dx}) and $V_{x}(t)$\ (\ref{Va}%
)\ provide details of the transport process, while their asymptotic values $%
D_{x}^{\infty },$ $V_{x}^{\infty }$\ represent the diffusion coefficient and
the pinch velocity that determine the impurity flux at the transport
space-time scale. \ \ 

\section{Effects of the parallel acceleration}

Typical results obtained for W ions with $Z=40$ in a turbulent plasma with
the parameters $\Phi =0.03,$\ $V_{d}=1,$\ $\lambda _{x}=5,$ $\lambda _{y}=2,$%
\ $\lambda _{z}=1,$\ $k_{0}=1$, $\tau _{d}=\infty $\ and $a/\rho _{i}=500$
are presented Figure 1. The time-dependent diffusion coefficient $D_{x}(t)$\
and the radial average velocity $V_{x}(t)$ are shown (solid lines) compared
to case of deuterium (D) ions (dashed-dotted lines) and to W ions with $%
a_{z}=0$ (dashed lines).

\begin{figure}[tbh]
\centerline{\includegraphics[height=5.3cm]{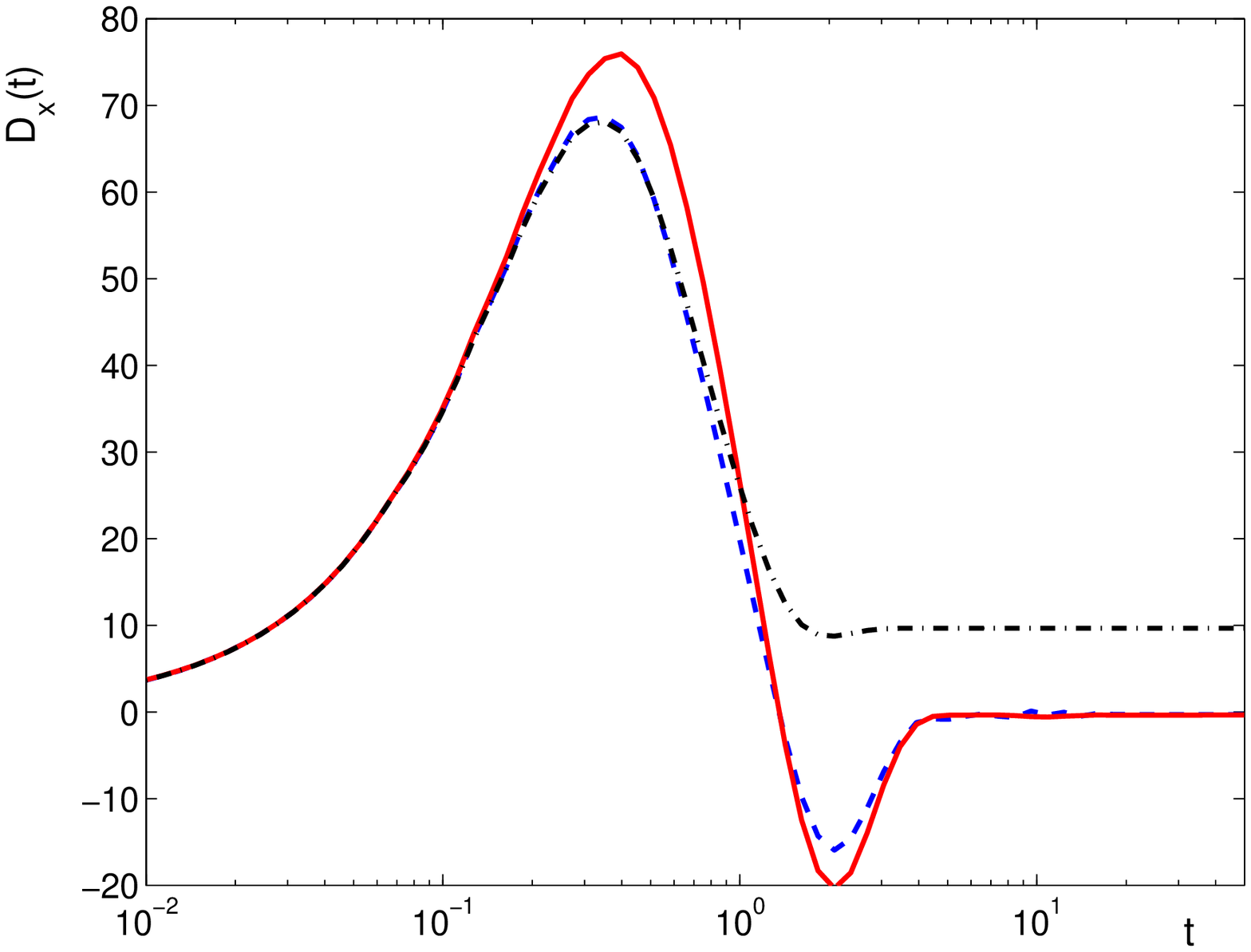} 
\includegraphics[height=5.3cm]{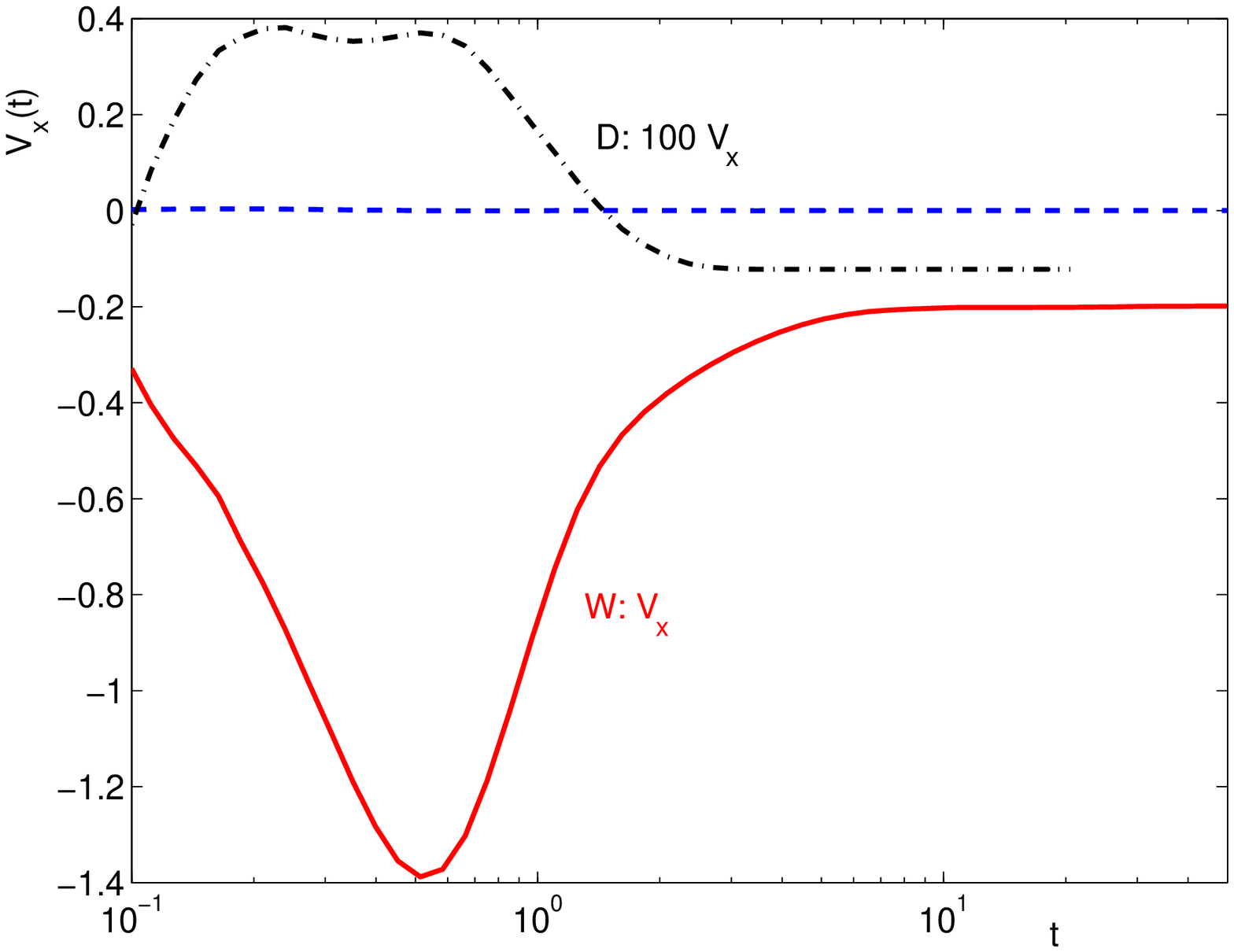}}
\caption{Typical results for W and D ions: the time dependent diffusion
coefficient $D_{x}(t) $ (left panel) and the average radial velocity $%
V_{x}(t)$ (right panel). The continuous (red) lines are for W ions with
parallel acceleration, the dashed (blue) lines are for the W ions with $%
a_z=0 $ and the dashed-dotted (black) lines are for D ions with parallel
acceleration.}
\label{Figure 1}
\end{figure}

The diffusion coefficients shown in Fig. \ref{Figure 1} (left panel) have a
similar time dependence for all three cases. The increase at small times
corresponds to the quasilinear regime that is identical for all examples. It
lasts for $t\ll \tau _{fl},$ where $\tau _{fl}$\ is the time of flight
defined as the ratio of $\lambda _{x}$\ and the amplitude of the stochastic
radial velocity, which in this units is $\tau _{fl}=\lambda _{x}\lambda
_{y}/K_{\ast }.$ The maximum of $D_{x}(t)$ appears at $\tau _{fl}=0.66$\
and, at larger times, the eddying (trapping) determines the decay of $%
D_{x}(t)$ that lasts until the decorrelation of the trajectories from the
potential produces the saturation. One can see that the saturation is at a
much smaller time for the D ions than for the W ions, which means that the
parallel decorrelation time $\tau _{z}^{\infty }$ is much smaller in the
first case. The parallel acceleration does not change the result for the D
ions. The same result is obtained with/without $a_{z}$ (the dashed-dotted
line). The parallel acceleration determines a modification of $D_{x}(t)$ for
the W ions (see the solid curve compared to the dashed one). It essentially
consists, in this case, of a small increase of the time of flight.

The main effect of the parallel acceleration is the generation of a radial
pinch. As seen in Fig. \ref{Figure 1} (right panel), a negative (inward)
average velocity $V_{x}(t)$ appears due to $a_{z}$. It has a transitory
large increase until $t\sim \tau _{fl},$ \ then it decays and eventually
saturates due to the parallel decorrelation. When $a_{z}$\ is neglected, $%
V_{x}(t)=0$ at any time. The radial pinch is much smaller for D ions than
for the W ions. As seen in the figure, only multiplied by $100$ the D pinch
velocity (dashed-dotted curve) reaches values comparable to the W pinch
velocity (solid curve).

These much larger effects of the parallel acceleration on the heavy impurity
transport compared to the case of D ions are rather surprising, because the
normalized parallel acceleration in Eq. (\ref{eqz1}) scales as $%
dv_{z}/dt\sim P_{a}\sim Z/\sqrt{A}.$\ It is smaller by a factor $0.3$ for W
compared to D ions, which strongly enables to predict very small effects on
heavy impurity transport.

The interaction of the parallel motion with the parallel transport is the
effect of the finite parallel correlation length $\lambda _{z},$ which makes
the EC (\ref{ECdef})\ a $z-$dependent function that decays with the increase
of $z.$\ The average of the EC over the parallel motion $z(t),$ solution of
Eq. (\ref{eqz1}), yields a time decaying function. This parallel
decorrelation process has the characteristic time $\tau _{z}^{\infty },$\
which is the asymptotic value of the effective parallel time defined in Eq. (%
\ref{tauef}). The values of $\tau _{z}^{\infty }$ and its scaling with the
parameters of the parallel motion are different for the D and W ions, as
shown below.

The variation range of the parallel velocity results from the energy
conservation%
\begin{equation}
v_{z}=\pm \sqrt{2\left( W-Z\Phi \phi \right) },  \label{vz}
\end{equation}
while its dynamics, reflected in the variation time, is determined by the
parallel acceleration $a_{z}.$

At $Z=1,$\ the potential energy is small, $\Phi \ll W,$\ and the velocity
can be approximated by $v_{z}\cong \pm \sqrt{2W}\left( 1-\Phi /2W\right) .$\
The acceleration determines for the D ions only a small fluctuation of $%
v_{z}(t)$ around the effective values $v_{z}^{eff}\cong \pm \sqrt{2W}.$ The
parallel displacements are $z=\pm $\ $\sqrt{2W/A}t,$\ and the decorrelation
time is approximated by $\tau _{z}^{\infty }\cong \lambda _{z}/\sqrt{W}$ 
\cite{Vlad02}.

Thus, the parallel decorrelation time is practically not modified by $a_{z}$%
\ at small $Z,$ because $v_{z}^{eff}$ does not depend on $\Phi $\ and $%
\lambda _{z}.$

At large $Z,$\ the potential energy is large, of the order of the total
energy $W$, $Z\Phi \sim W.$\ In these conditions, the trajectories cannot
reach the regions with large, positive $\phi $\ and the Lagrangian potential
has an upper limit%
\begin{equation}
\phi (\mathbf{x}(t),z(t))<\phi _{\max }=\frac{W}{Z\Phi }.  \label{fim}
\end{equation}%
In addition, the kinetic energy is larger than the total energy\ $W$ in the
regions with negative potential. Both the average and the fluctuation
amplitude of $v_{z}(t)$\ are functions of $W,$ $Z$ and $\Phi .$ They also
depend on $\lambda _{z}$\ through the characteristic variation time of $%
v_{z}(t),$ which is determined by the acceleration $a_{z}\sim 1/\lambda
_{z}. $ The analytical estimation of $v_{z}^{eff}(W,Z,\Phi ,\lambda _{z})$
and $\tau _{z}^{\infty }(W,Z,\Phi ,\lambda _{z})$\ is not possible in this
case, but only the general behaviour with the parameters of the parallel
motion. The range of variation of $\left\vert v_{z}(t)\right\vert $ is the
interval $\left[ 0,~W+Z\Phi \right) ,$ where the lower limit is determined
by $\phi _{\max }$\ in Eq. (\ref{fim})\ and the upper limit corresponds to
the amplitude of the order $-\Phi $ of the negative potential. This shows
that $v_{z}^{eff}$ increases with $W,$ $Z$ and $\Phi ,$ and that $%
v_{z}^{eff}(W,Z,\Phi ,\lambda _{z})>\sqrt{2W}.$\ The dynamics of $v_{z}(t)$
that is determined by $a_{z}\sim P_{a}/\lambda _{z}$\ leads to the decrease
of $v_{z}^{eff}$\ at the increase of $\lambda _{z}.$

Thus, the parallel decorrelation time is modified by $a_{z}$\ at large $Z,$
and depends on all the parameters of the parallel motion. It can be
approximated by 
\begin{equation}
\tau _{z}^{\infty }(W,Z,\Phi ,\lambda _{z})\cong \lambda _{z}\sqrt{A}%
/v_{z}^{eff},  \label{tauz}
\end{equation}
which is a decreasing function of $W,$ $Z,$ $\Phi $\ and an increasing
function of $\lambda _{z}$ (as $\lambda _{z}^{\alpha }$ with $\alpha >1).$

The influence of the decorrelation time on the diffusion is different in the
quasilinear ($\tau _{z}^{\infty }<\tau _{fl})$ and the trapping ($\tau
_{z}^{\infty }>\tau _{fl})$ regimes. The asymptotic diffusion coefficient
scales as%
\begin{equation}
D_{x}^{\infty }\sim \left\{ 
\begin{array}{c}
\Phi ^{2}\tau _{z}^{\infty },\ \ \ \ \tau _{z}^{\infty }<\tau _{fl} \\ 
\Phi ^{\gamma }\tau _{z}^{\gamma -1},\ \ \tau _{z}^{\infty }>\tau _{fl}%
\end{array}%
\right. ,  \label{Dscal}
\end{equation}%
\ where $0<\gamma <1$.\ \ 

The effects of the interaction of the parallel motion with the perpendicular
transport through the parallel decorrelation explain the results obtained
for the diffusion coefficients. The influence of the parallel acceleration
is negligible for the D ions and noticeable for W ions. The strongest
difference appears due to the dependence of $\tau _{z}^{\infty }$ on the
mass number, which leads for the case presented in Figure 1 (left panel) to $%
\tau _{z}^{\infty }=1.15$ for the D ions (dashed-dotted line) and $\tau
_{z}^{\infty }=11.25$ for the W ions (dashed line). The diffusion is in the
trapping regime in both cases, which corresponds, according to Eq. (\ref%
{Dscal}), to much smaller $D_{x}^{\infty }$ for the W ions compared to D
ions, as seen in Figure 1 (left panel).

The the radial pinch velocity seen in Figure 1 (right panel) cannot be
explained by the parallel decorrelation process. A different interaction
process provides the physical mechanism of pinch generation, as demonstrated
in the next section.

\section{The pinch mechanism}

Particle trajectories described by Eqs. (\ref{eqm})-(\ref{eqmz}) with $%
\lambda _{z}\rightarrow \infty $ (two-dimensional potentials) have both
stochastic and quasi-coherent aspects. The coherent motion is determined by
the trapping or eddying in the structure of the potential, which determine
small structures that produce a micro-confinement process \cite{VS17}. It
hinders the diffusive transport by decreasing the diffusion coefficient. The
quasi-coherent component of the motion can also yield flows \cite{Vlad06}-%
\cite{Vlad18}. We have shown \cite{VS18} that a special quasi-coherent
effect, that is neither structure nor flow, appears in the stochastic
transport in the presence of an average poloidal velocity $V_{p}$. It
consists of two average radial velocities in opposite directions, which
exactly compensate. This pair of drifts are named in \cite{VS18} hidden
drifts (HDs) because they do not yield an average velocity in these
conditions.

The HDs are essentially determined by the existence of average displacements
of the trajectories that start from same values of the potential $\phi ^{0}$%
, and by the special property of these conditional averages $\left\langle 
\mathbf{x}(t)\right\rangle _{\phi ^{0}}$ of having the sign correlated to
the sign of $\phi ^{0}.$ These quantities, evaluated by DTM from Eq. (\ref%
{xme}), are\ 

\begin{equation}
\left\langle x(t)\right\rangle _{\phi ^{0}}=\int d\phi _{x}^{0}d\phi
_{y}^{0}d\phi _{z}^{0}~P(\phi ^{0},\phi _{i}^{0})X(t;\phi ^{0},\phi
_{i}^{0}).  \label{x-fi}
\end{equation}

\begin{figure}[tbh]
\centerline{\includegraphics[height=5.2cm]{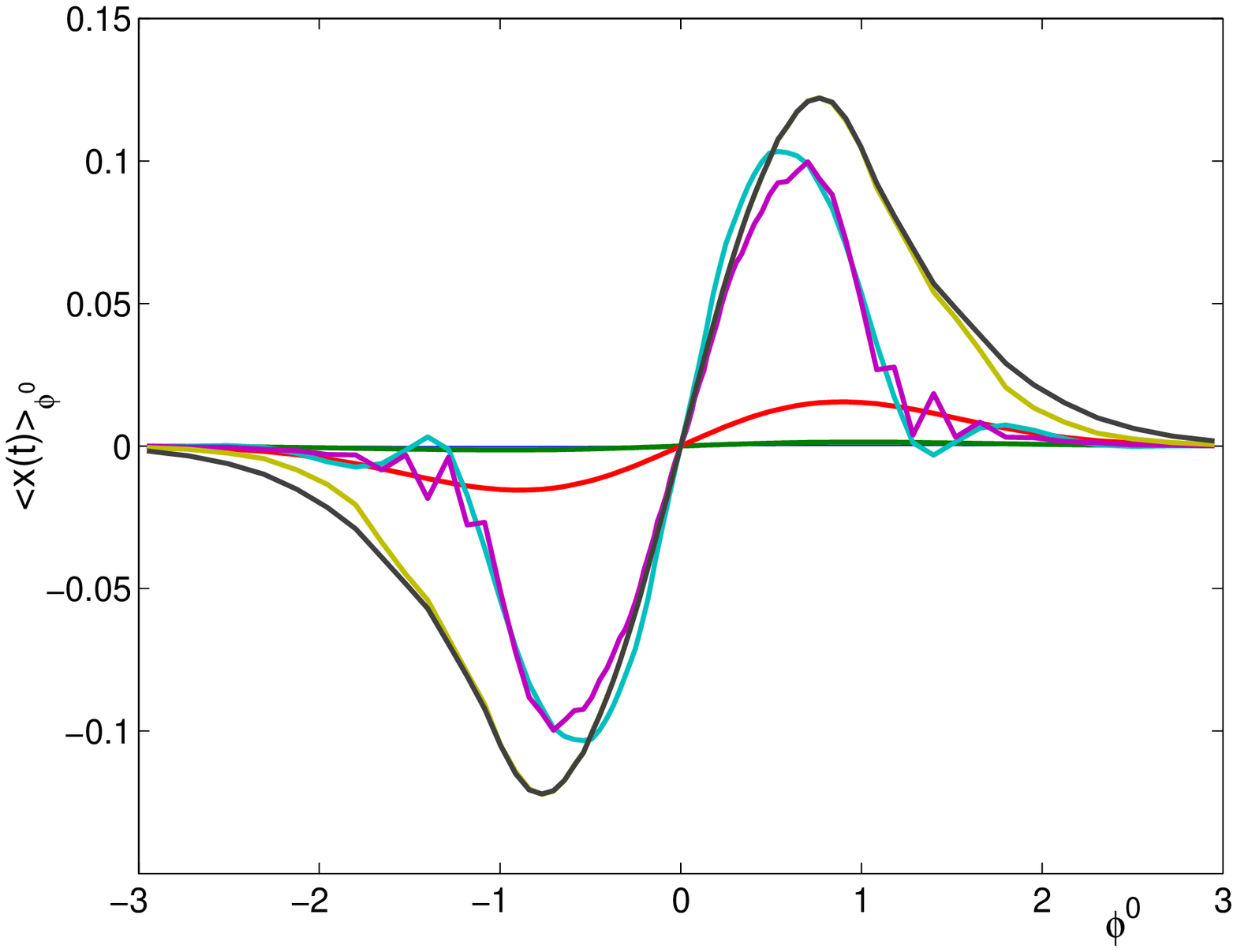} 
\includegraphics[height=5.2cm]{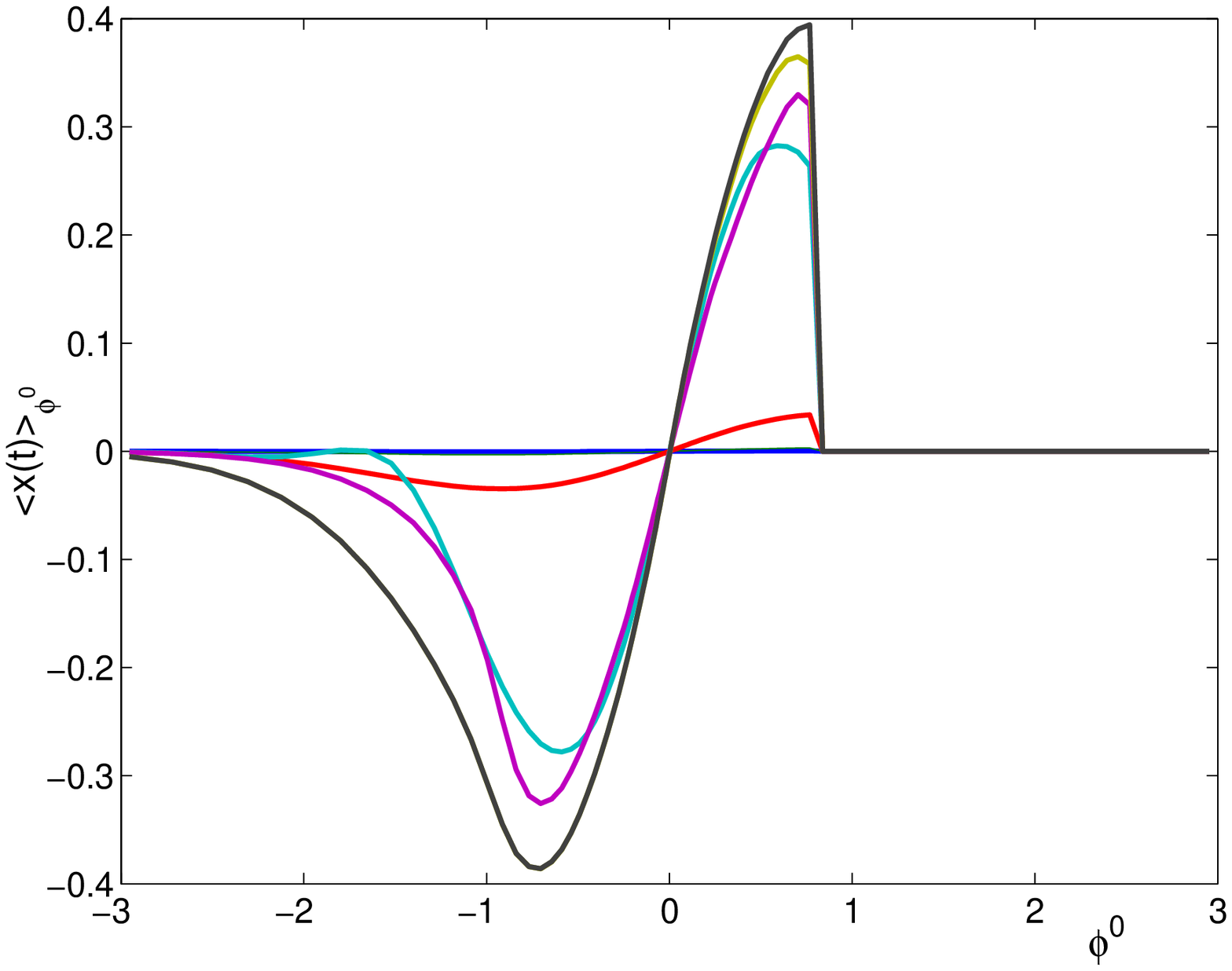}}
\caption{The average displacements conditioned by the initial potential $%
\protect\phi ^{0}$ as functions of $\protect\phi ^{0}$ for several times.
The parallel acceleration is neglected in the left panel and is considered
in the right panel. }
\label{Figure 2}
\end{figure}

The conditional displacements are zero in the case of the motion determined
only by the electric drift, but they have finite values in the presence of
an average poloidal velocity $V_{p}$. In the absence of the parallel
acceleration (two-dimensional potentials), $\left\langle \mathbf{x}%
(t)\right\rangle _{\phi ^{0}}$\ is an anti-symmetrical function of $\phi
^{0} $\ and it leads, by integration over $\phi ^{0},$\ to zero average
displacement. A typical example is shown in Fig. \ref{Figure 2} (left
panel), which corresponds to the transport of the W ions with $a_{z}=0$
represented in Fig. \ref{Figure 1} by the dashed lines.

This special type of quasi-coherent motion is generated by the average
poloidal velocity $V_{p},$ which is equivalent with an average potential
that adds to the stochastic potential. The contour lines of the total
potential $\phi _{t}(\mathbf{x)}=\phi (\mathbf{x)}+xV_{p}$\ show a complex
structure having strips of open lines that oscillate between islands of
closed lines. The trajectories are along the contour lines of the potential,
and thus they are of two types: trapped (closed) and free (with unlimited
displacements along $\mathbf{V}_{p})$. The probability of finding the
trajectory along a closed contour line is larger on the side on which $%
\mathbf{V}_{p}$ is opposite to the stochastic velocity than on the other
side. This leads to average radial displacements on the contour line $\phi
^{0}$ that are positive for $\phi ^{0}>0$ and negative for $\phi ^{0}<0.$
The free trajectories also contribute to the ordered conditional
displacements. The Lagrangian invariance of $\phi _{t}(\mathbf{x}(t)\mathbf{)%
}=\phi (\mathbf{x}(t)\mathbf{)}+x(t)V_{p}=\phi ^{0}$\ constrains these
trajectories to oscillate around the line $x=\phi ^{0}/V_{p},$ which is the
average for the trajectories that start from $\phi ^{0}.$

An important property can be deduced from the physical image of the
quasi-coherent process generated by the poloidal average velocity. The
change of the sign of $V_{p}$ determines the change of the sign of the
conditional average displacement $\left\langle \mathbf{x}(t)\right\rangle
_{\phi ^{0}}.$

The average displacements conditioned by the sign of the potential 
\begin{equation}
\left\langle x(t)\right\rangle _{+}=\int_{0}^{\infty }d\phi ^{0}\left\langle
x(t)\right\rangle _{\phi ^{0}},\ \ \left\langle x(t)\right\rangle
_{-}=\int_{-\infty }^{0}d\phi ^{0}\left\langle x(t)\right\rangle _{\phi ^{0}}
\label{x-sign}
\end{equation}%
determine, using Eq. (\ref{Va}), two opposite radial velocities $V_{+},$ $%
V_{-}$ that exactly compensate $V_{+}+V_{-}=0$ due to the anti-symmetry of $%
\left\langle x(t)\right\rangle _{\phi ^{0}}$ with respect to the initial
potential $\phi ^{0}$ (see \cite{VS18}, \cite{VS18-2} for details).

The HDs represent a reservoir for direct transport, because perturbations
produced by other components of the motion can affect the equilibrium of the
HDs leading to an average velocity. We have shown \cite{VS18-2} that the
polarization drift determines a significant modification of the symmetry of
the HDs and provides a mechanism for radial pinch generation. Essentially,
this pinch appears due to the compressibility effect of the polarization
drift.

We show here that the three-dimensional stochastic motion described by Eqs. (%
\ref{eqm})-(\ref{eqmz}) can influence the equilibrium of the HDs.

An important difference between the three-dimensional and two-dimensional
motion is that the Lagrangian potential is not invariant. This means that
the trajectories in the subensemble S do not evolve on the contour lines \ $%
\phi (\mathbf{x})=\phi ^{0},$ but they move up and down $\phi (\mathbf{x},z)$
according to the parallel acceleration and velocity. As a consequence, the
conditional averages $\left\langle x(t)\right\rangle _{\phi ^{0}}$\ undergo
a complex averaging process that influences their anti-symmetrical
dependence on $\phi ^{0}.$\ The parallel acceleration moves the trajectories
toward the minima of the stochastic potential, which favours the $%
\left\langle x(t)\right\rangle _{\phi ^{0}}$\ with $\phi ^{0}<0.$ On the
other hand, the parallel velocity increases in the regions with negative
potential and decreases in the regions with positive potential. The ions
spend smaller time at negative than at positive potential, which partly
compensates the attraction towards the potential minima, and yields a small
perturbation of the symmetry of the HDs.

This explains the small pinch velocity observed for D ions.

The heavy ions with large ionization rates $Z$ have smaller acceleration,
but also a much higher potential energy (larger than for D ions by the
factor $Z).$\ Then, even for small amplitudes of the\ turbulence ($\Phi $ of
the order $10^{-2}),$ the trajectories cannot reach the maxima of the
stochastic potential, because the invariance of the energy (\ref{W}) imposes 
$W-Z\Phi \phi ^{0}=v_{z}^{2}/2>0.$ The result is the cut of $\left\langle
x(t)\right\rangle _{\phi ^{0}}$ at large $\phi ^{0},$\ as seen in Figure 2.b
(obtained with DTM for the case presented by the solid lines in Figure 1). A
strong symmetry breaking of the conditional displacements is produced at
large $Z,$ which yields, using Eq. (\ref{Va}), an average radial velocity $%
V_{x}(t).$\ The maximum allowed potential $\phi _{\max }$ Eq. (\ref{fim})
decreases with the increase of $Z,$ which leads to the increase of the
average displacement and of the pinch velocity $V_{x}^{\infty }$.

This explains the significant pinch velocity produced by the parallel
acceleration for W ions.\ 

\bigskip

\begin{figure}[tbh]
\centerline{
\includegraphics[height=5.2cm]{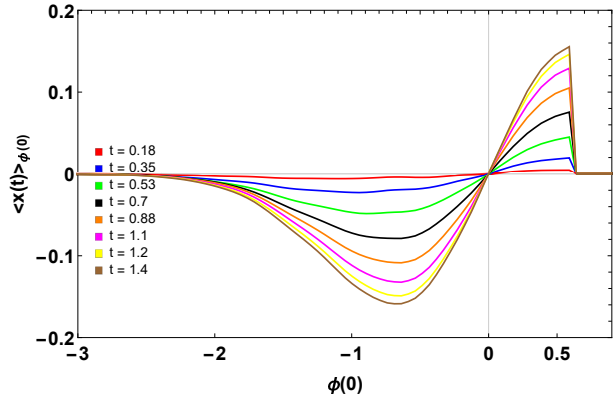}}
\caption{Validation of the pinch mechanism by DNS. The conditional
displacements $\left\langle x(t)\right\rangle _{\protect\phi ^{0}}$ obtained
with DNS. }
\label{Figure 3hd}
\end{figure}

The physical mechanism for the generation of the radial pinch is validated
using DNS. Fig. \ref{Figure 3hd} confirms the existence of the conditional
displacements $\left\langle x(t)\right\rangle _{\phi ^{0}}$\ and the
perturbation produced by the parallel acceleration, which essentially
consists of forbidding the trajectories to reach the maxima of the
potential. It yields average displacements as function of the initial value
of the potential that are similar to those obtained by DTM (shown in Fig. %
\ref{Figure 2} (right panel)).

\begin{figure}[tbh]
\centerline{\includegraphics[height=5.2cm]{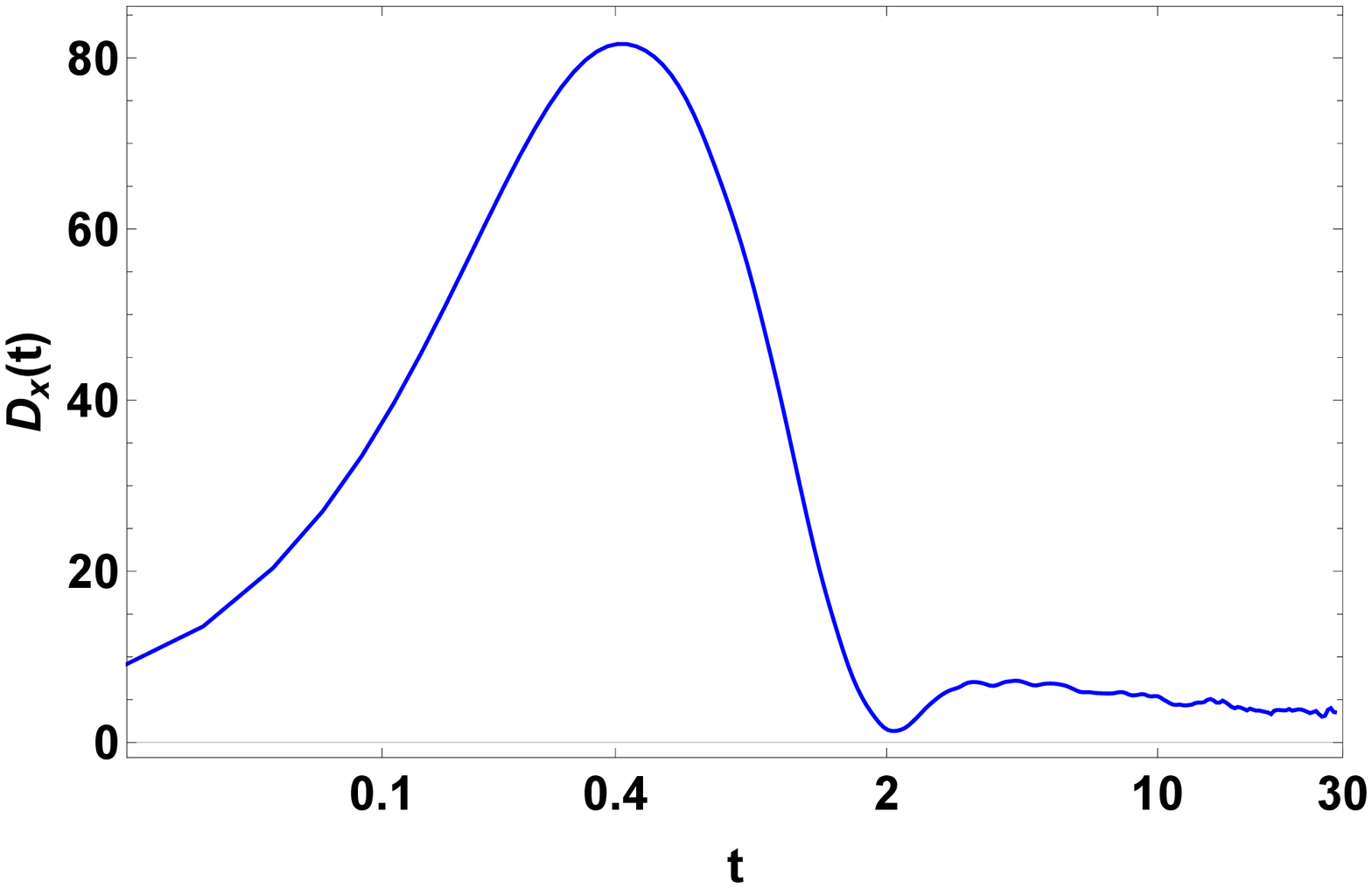} 
\includegraphics[height=5.2cm]{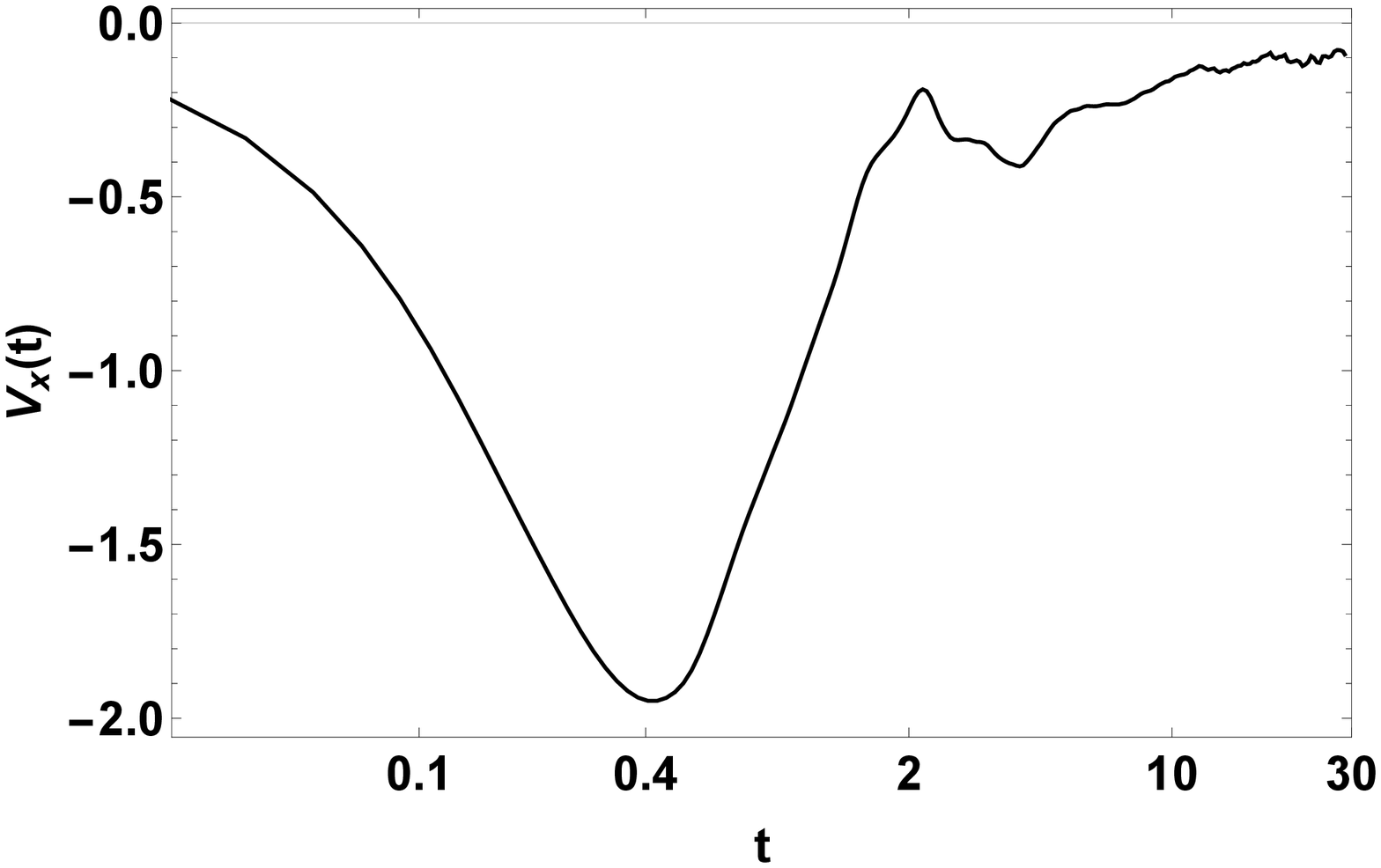}}
\caption{Validation of the pinch mechanism by DNS. The time dependent
diffusion coefficient $D_{x}(t)$ (left panel) and the average radial
velocity $V_{x}(t)$ (right panel). }
\label{Figure 3}
\end{figure}

The time dependent pinch velocity and diffusion coefficient obtained from
the numerical simulation of the stochastic trajectories are presented in
Fig. \ref{Figure 3}. They correspond to the W ion transport for the set of
parameters mentioned at the beginning of this Section, which yield using DTM
the results shown in Fig. \ref{Figure 1} (continuous lines). One can see
that the results of the simulation are similar to those of the DTM for both $%
V_{x}(t)$ and $D_{x}(t).$\ This shows that DTM is qualitatively adequate for
the study of the three-dimensional model (\ref{eqm})-(\ref{eqmz}). This
conclusion is in agreement with previous studies \cite{Jenko}.

There are, however, important differences that result from the approximation
used in the DTM. They consists of the overestimation of trajectory trapping,
which leads to smaller $D_{x}^{\infty }$ and larger $V_{x}^{\infty }$. The
overestimation of the pinch velocity is stronger in the case of time
dependent potentials. The results presented in the next Section are obtained
using DNS.

\bigskip

\section{Characterization of the pinch velocity}

The impurity ion transport described by Eqs. (\ref{eqm})-(\ref{eqmz})
depends on a large number of parameters. Taking the case of W ions in large
size plasmas of ITER type, we fix $A=184$ and $\rho _{\ast }=\rho
_{i}/a=1/500,$ and remain with $9$ dimensionless parameters: $K_{\ast },$ $%
\lambda _{z}/a,$ $V_{p}/V_{\ast },$ $P_{a},$ $\tau _{d}/\tau _{0},$ $W,$ $%
\lambda _{x}/\rho _{i},$ $\lambda _{y}/\rho _{i}$ and $k_{0}\rho _{i}.$

The first three parameters are essential because they describe the main
ingredients of the pinch mechanism: a turbulent state of plasma ($K_{\ast })$
with three-dimensional stochastic potential (finite $\lambda _{z})$ and an
average poloidal velocity ($V_{p}).$ The characteristics of the impurity
ions\ are represented by $P_{a}.$ The time variation of the stochastic
potential is expected to damage the ordered component of the motion and to
favour the random aspects. It is thus essential to investigate the
dependence of W transport on the decorrelation time $\tau _{d}.$ The
dependence of the pinch velocity on the energy of the ions $W$ could be
important as a control method. The other three parameters describe details
of the shape of the turbulence EC (\ref{EC}), which are less important for
the pinch mechanism produced by the parallel acceleration.

$K_{\ast }=\Phi /\rho _{\ast }$ is the measure of turbulence amplitude $\Phi
=eA_{\phi }/T_{i}$. The latter is also contained in the parameter of the
parallel acceleration $P_{a}$. For a more clear presentation of the results,
we analyze the dependence of the W ion transport on the physical parameters $%
\Phi $ and $Z$ rather than on $K_{\ast }$ and $P_{a}.$

The physical range of the parameters is explored around a basic case with $%
\Phi =0.03,$ $Z=40$ (corresponding to $K_{\ast }=15,$ $P_{a}=0.09),$ $%
\lambda _{z}=0.5,$ $V_{d}=1,~\tau _{d}=\infty $ (static potential), $W=1,$ $%
\ \lambda _{x}=5,$ $\lambda _{y}=2,$ and $k_{0}=1.$ The units in the figures
are $\rho _{i}V_{\ast }$ for the diffusion coefficient, $\rho _{i}$ for the
average trajectories, $V_{\ast }$ for the pinch velocity and $\tau _{0}$ for
the time.

The analysis is performed using numerical simulations (the DNS methods and
codes described in Section 3.1). \ \ \ \ \ 

\begin{figure}[tbh]
\centerline{\includegraphics[height=5.2cm]{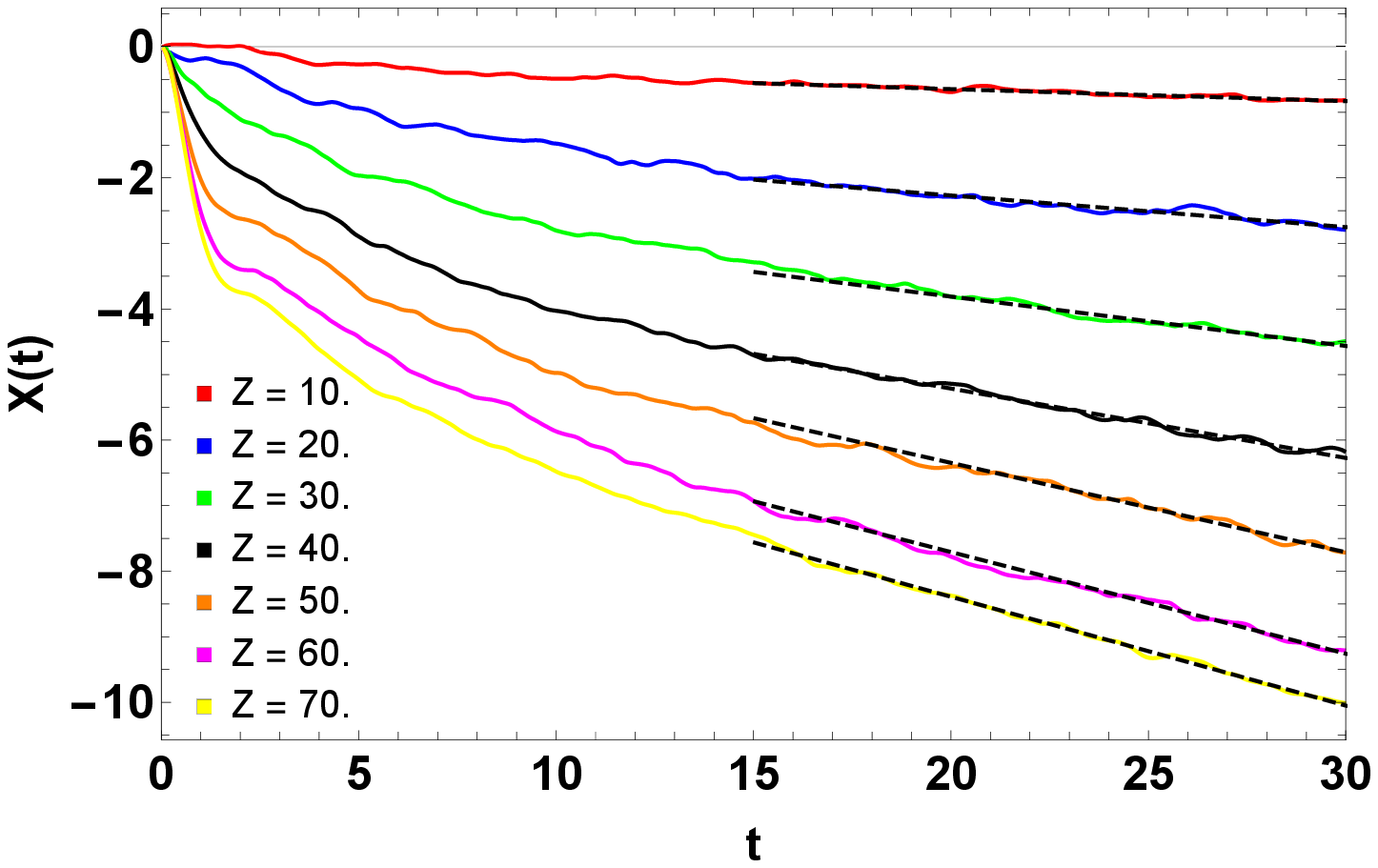}
\includegraphics[height=5.2cm]{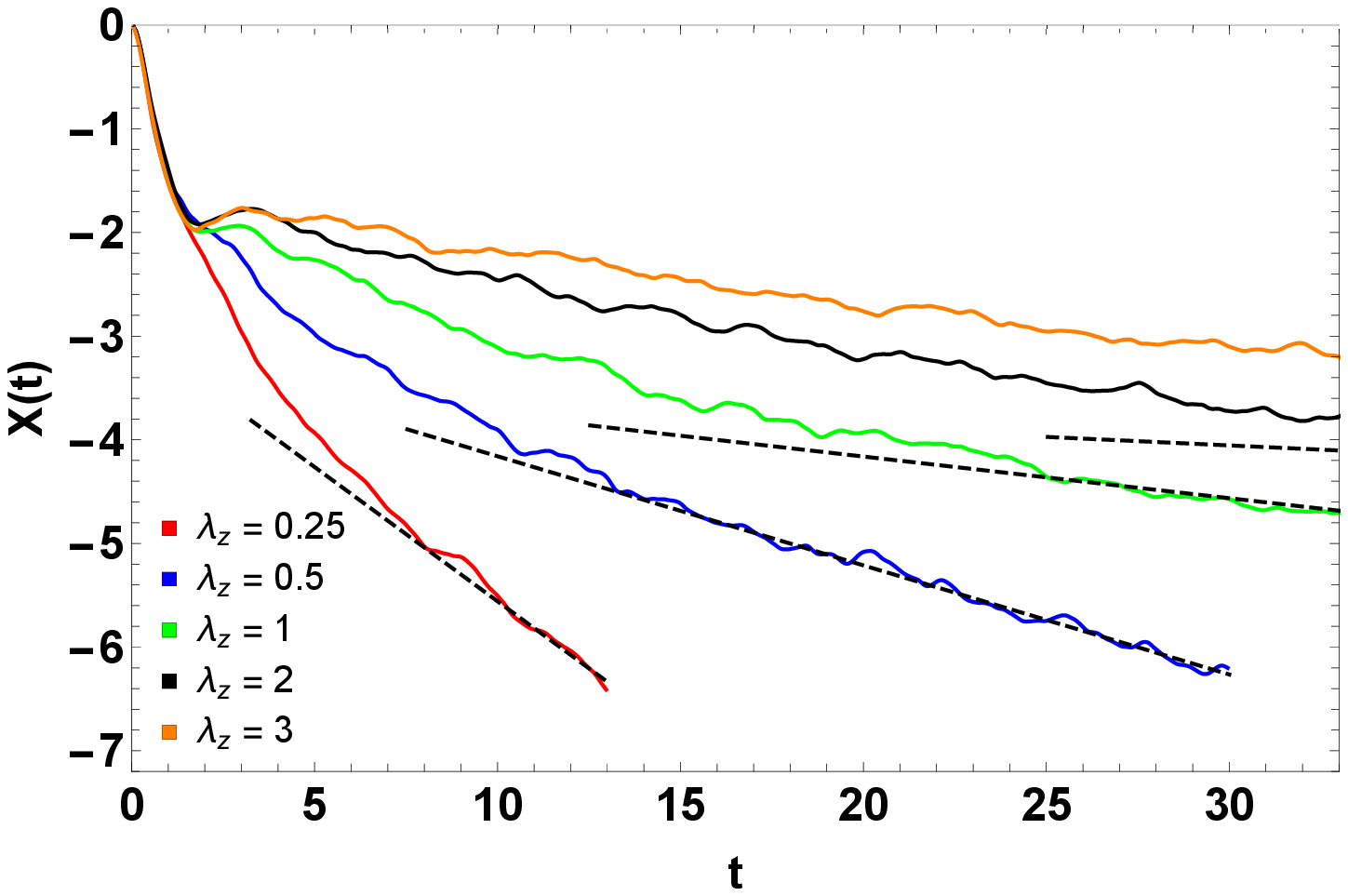}}
\caption{The average radial dispacement of the W ions for several values of $%
Z$ (left panel) and of $\protect\lambda_{z}$ (right panel). The other
parameters correspond to the basic case. }
\label{xmed}
\end{figure}
\bigskip

\begin{figure}[tbh]
\centerline{\includegraphics[height=5.2cm]{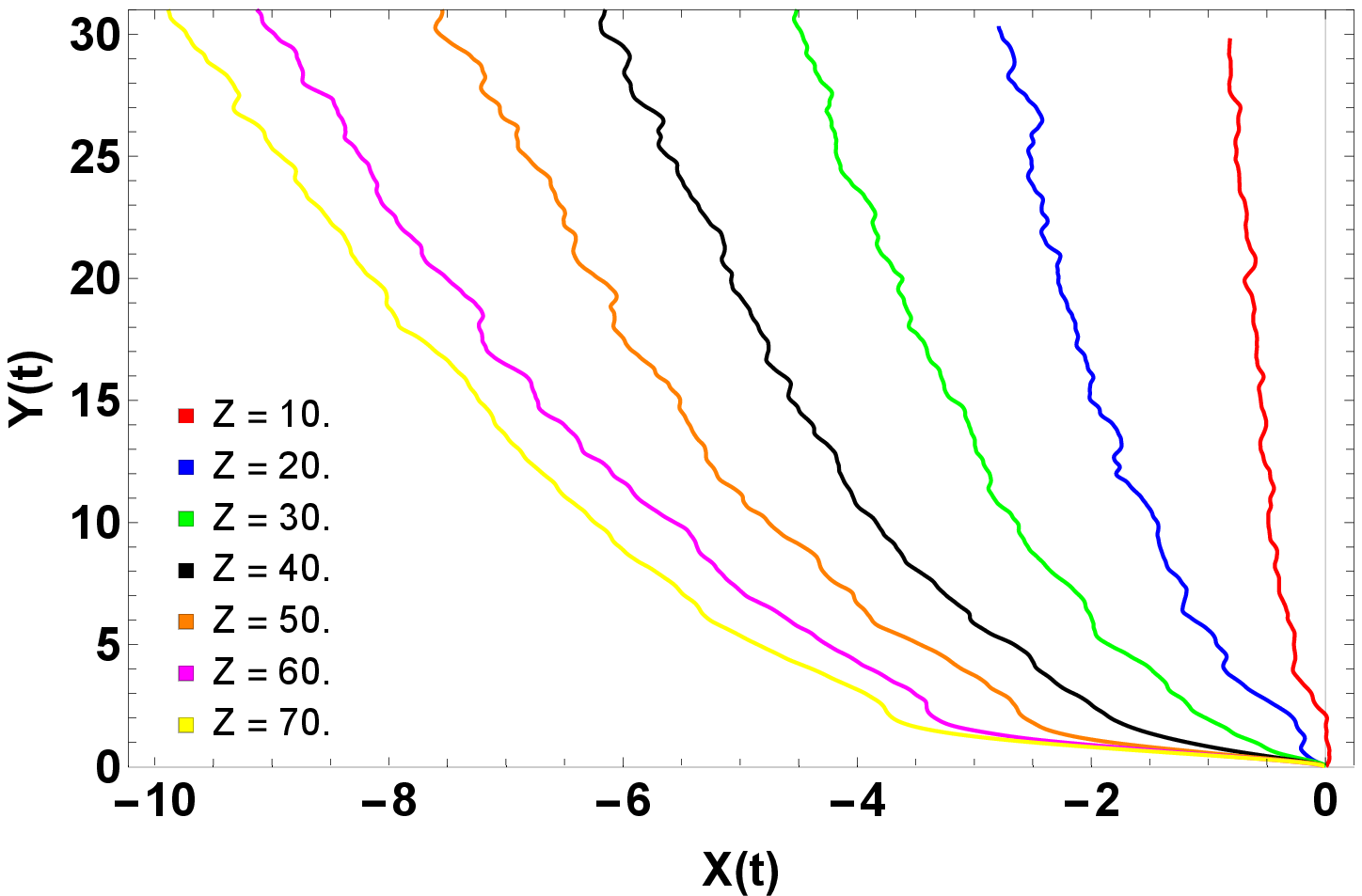}
\includegraphics[height=5.2cm]{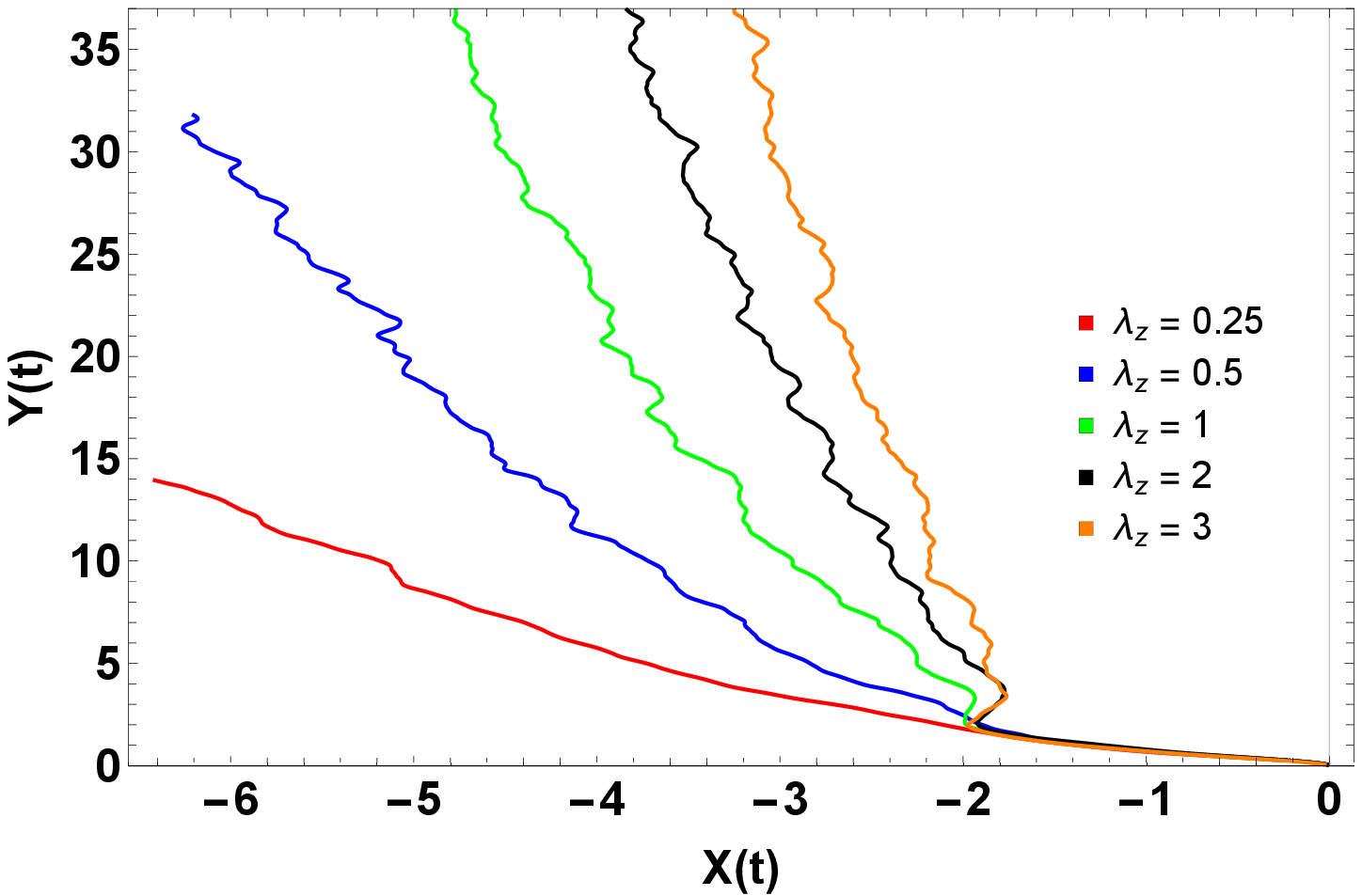}}
\caption{The average paths of the W ions for several values of $Z$ (left
panel) and of $\protect\lambda_{z}$ (right panel). The other parameters
correspond to the basic case. }
\label{XY}
\end{figure}
\bigskip

The existence of the radial pinch velocity can be clearly seen in Fig. \ref%
{xmed}, which presents examples of radial displacements $X(t)=\left\langle
x(t)\right\rangle $. A fast increase of $X(t)$ appears in all cases at small
times, followed by a transitory evolution (that depends on the parameters of
the process), which eventually leads to the asymptotic regime. The latter is
always linear in time and corresponds to the (asymptotic) radial pinch. We
note that the poloidal average velocity is not invariant as in
two-dimensional potentials. Starting from $V_{p}$, it has a transitory
variation that can end with a stabilized asymptotic slightly different of $%
V_{p}$. Examples of the paths of the average trajectories are presented in
Fig. \ref{XY}. They show that the average poloidal motion is not simply $%
Y(t)=V_{p}t,$ but it depends on the other parameters (especially on $\lambda
_{z}$, which controls the parallel acceleration).

We present below the results obtained for the dependence of the asymptotic
values $V_{x}^{\infty }$\ and $D_{x}^{\infty }$\ on each parameter. Scaling
laws are derived and physical explanations are deduced. The latter are based
on the two interaction mechanisms between the parallel motion and the
perpendicular transport: the symmetry breaking of the HDs that generates the
pinch and the parallel decorrelation that influences both $V_{x}^{\infty }$\
and $D_{x}^{\infty }.$\ The parallel decorrelation time $\tau _{z}^{\infty
}(W,Z,\Phi ,\lambda _{z}),$\ Eq. (\ref{tauz}), is a decreasing function of $%
W,$ $\Phi $ and $Z$ and an increasing function of $\lambda _{z},$ as
discussed in Section\ 4.

\begin{itemize}
\item Turbulence amplitude $\Phi $
\end{itemize}

The amplitude of the turbulence influences the electric drift velocity, the
parallel acceleration and the potential energy. It has a complex effect on
the pinch velocity and on the diffusion coefficient.

The increase of the electric drift determines the decay of the time of
flight as $\Phi ^{-1},$ and a stronger transient growth in the quasilinear
regime $\left( t<\tau _{fl}\right) $ for both $D_{x}(t)\sim \Phi ^{2}t^{2}$
and $V_{x}(t)\sim \Phi t$. Trajectory eddying combined with the increase of
the potential energy and of the parallel acceleration modifies the
dependence on $\Phi $ in the nonlinear regime, and, consequently, in the
asymptotic $V_{x}^{\infty }$ and $D_{x}^{\infty }$.

The asymptotic radial velocity $V_{x}^{\infty }$ is shown in Fig. \ref%
{Figure 4} (left panel) as function of $\Phi $. The pinch is negative
(inward) for the whole range of $\Phi $ and it increases with $\Phi .$ The
dependence is approximately linear for $\Phi \leq 0.04,$ and a tendency of
saturation can be observed at larger $\Phi .$\ 

The saturation of $V_{x}^{\infty }$ at large $\Phi $ is determined by the
energy conservation, which prevents the trajectories to reach the regions
with positive values of $\phi $ above the limit $\phi _{\max }$ defined in
Eq. (\ref{fim}). This determines the cut of $\left\langle x(t)\right\rangle
_{\phi ^{0}}$\ seen in Fig. \ref{Figure 2} (right panel), which \ destroys
the equilibrium of the HDs. The maximum perturbation of the HDs corresponds
to the limit $Z\Phi \rightarrow \infty ,$ which eliminates the whole
positive range of $\phi ^{0}$ ($\phi _{\max }\rightarrow 0)$. The pinch
velocity saturates for $Z\Phi \rightarrow \infty $ (practically for $Z\Phi
\gg W)$ at a value that equals the negative HD.

\begin{figure}[tbh]
\centerline{\includegraphics[height=5.2cm]{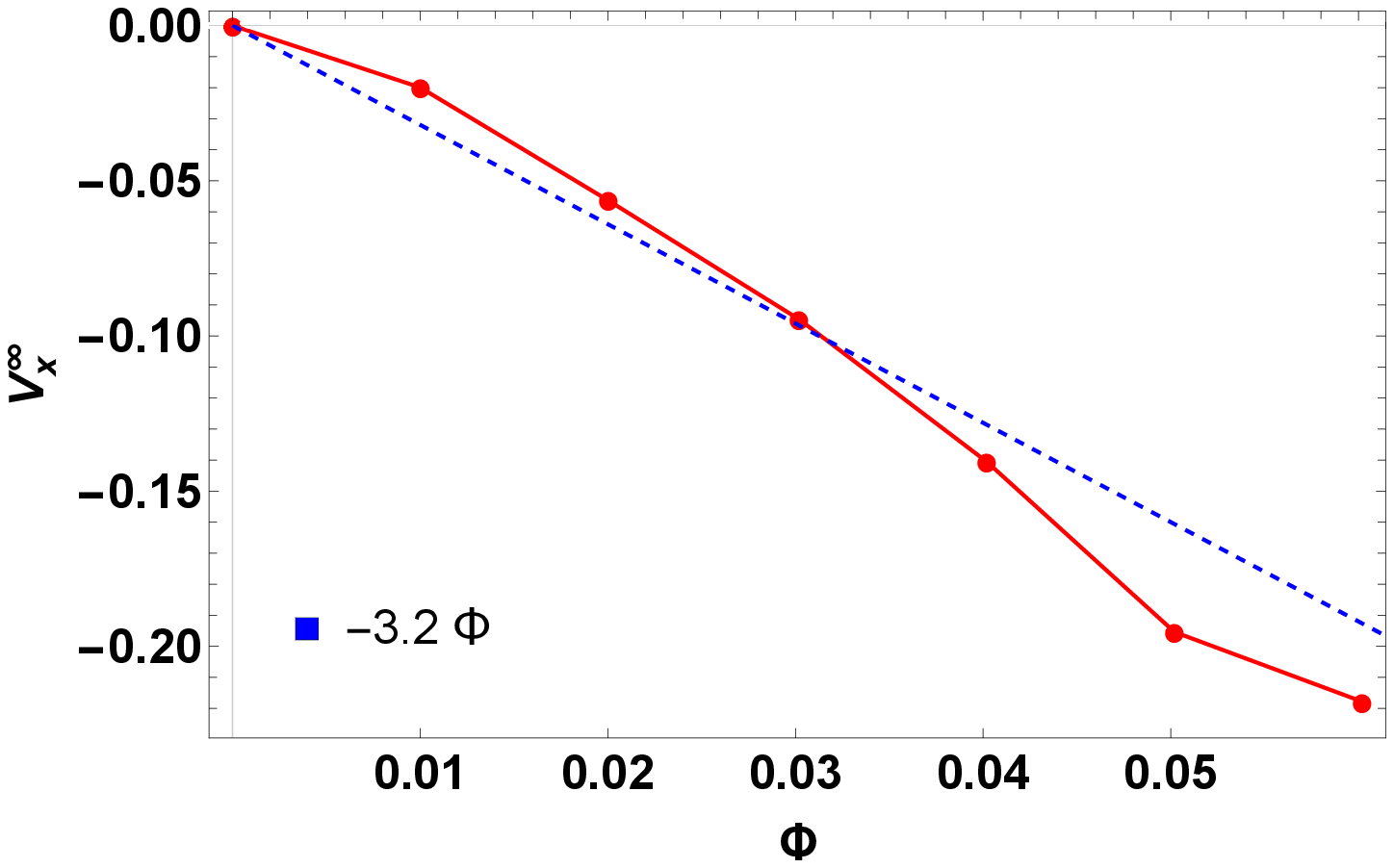} 
\includegraphics[height=5.2cm]{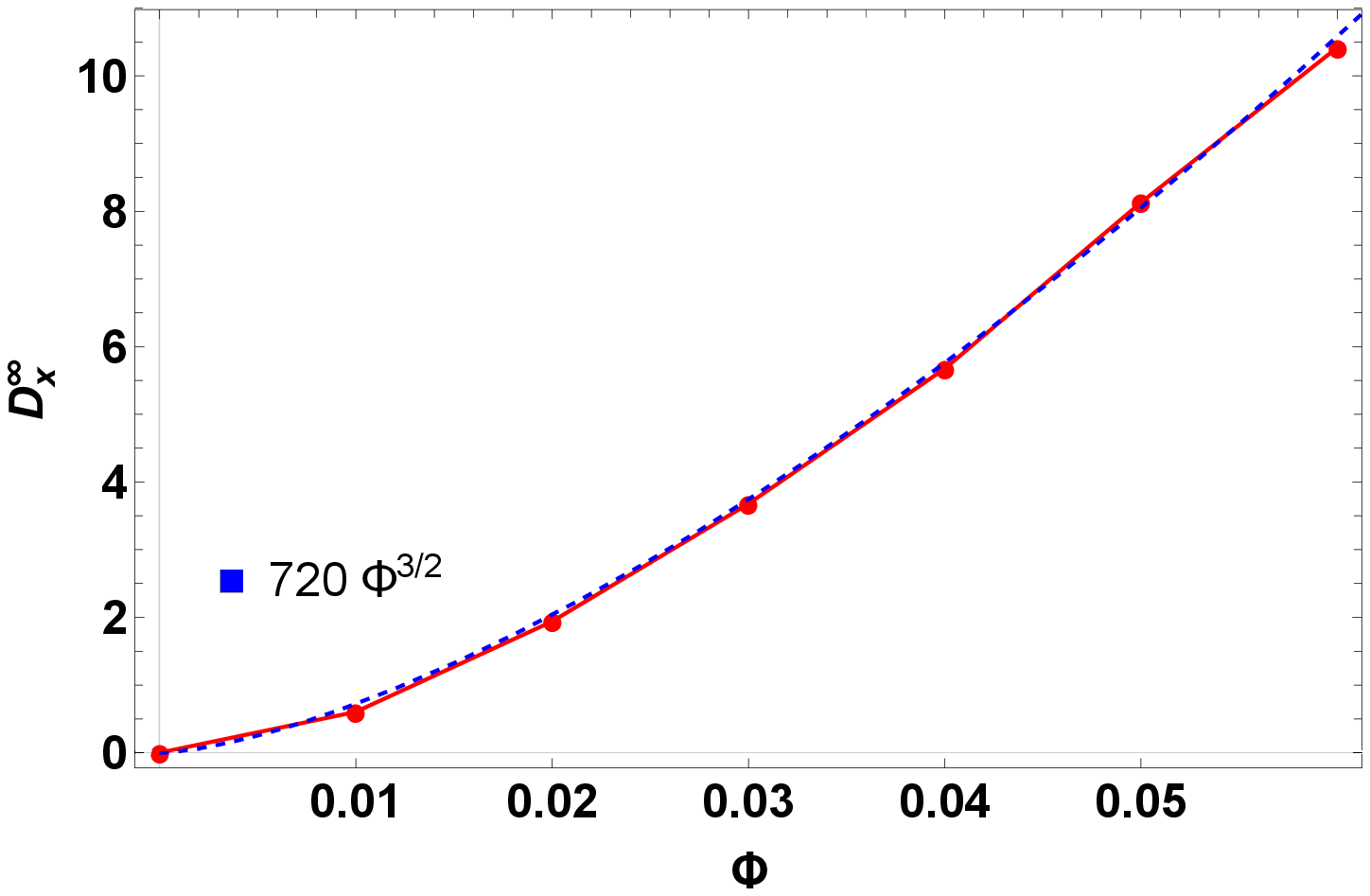}}
\caption{W transport dependence on turbulence amplitude : the asymptotic
pinch velocity (left panel) and the diffusion coefficient (right panel). }
\label{Figure 4}
\end{figure}

The asymptotic diffusion coefficient $D_{x}^{\infty }$ increases with the
increase of $\Phi $ according to the law $D_{x}^{\infty }\sim \Phi ^{\gamma
} $ with $\gamma =1.5,$ as seen in Fig. \ref{Figure 4} (right panel). The
values $1<\gamma <2$ define the super-Bohm regime. Such regime is unusual in
the presence of trajectory trapping or eddying, which yields the scaling (%
\ref{Dscal}) with $0<\gamma <1.$ This stronger increase of $D_{x}^{\infty }$
is the effect of the parallel acceleration through the effective
decorrelation time $\tau _{z}^{\infty }$. As discussed in Section 4, $\tau
_{z}^{\infty }$ is a decreasing function of $\Phi .$ The supplementary
dependence on $\Phi $ through $\tau _{z}^{\infty }(\Phi )$ increases the
exponent $\gamma .$ Thus, the super-Bohm regime is the result of trajectory
trapping coupled to the parallel accelerated motion.

\begin{itemize}
\item Ionization rate $Z$
\end{itemize}

\begin{figure}[tbh]
\centerline{\includegraphics[height=5.2cm]{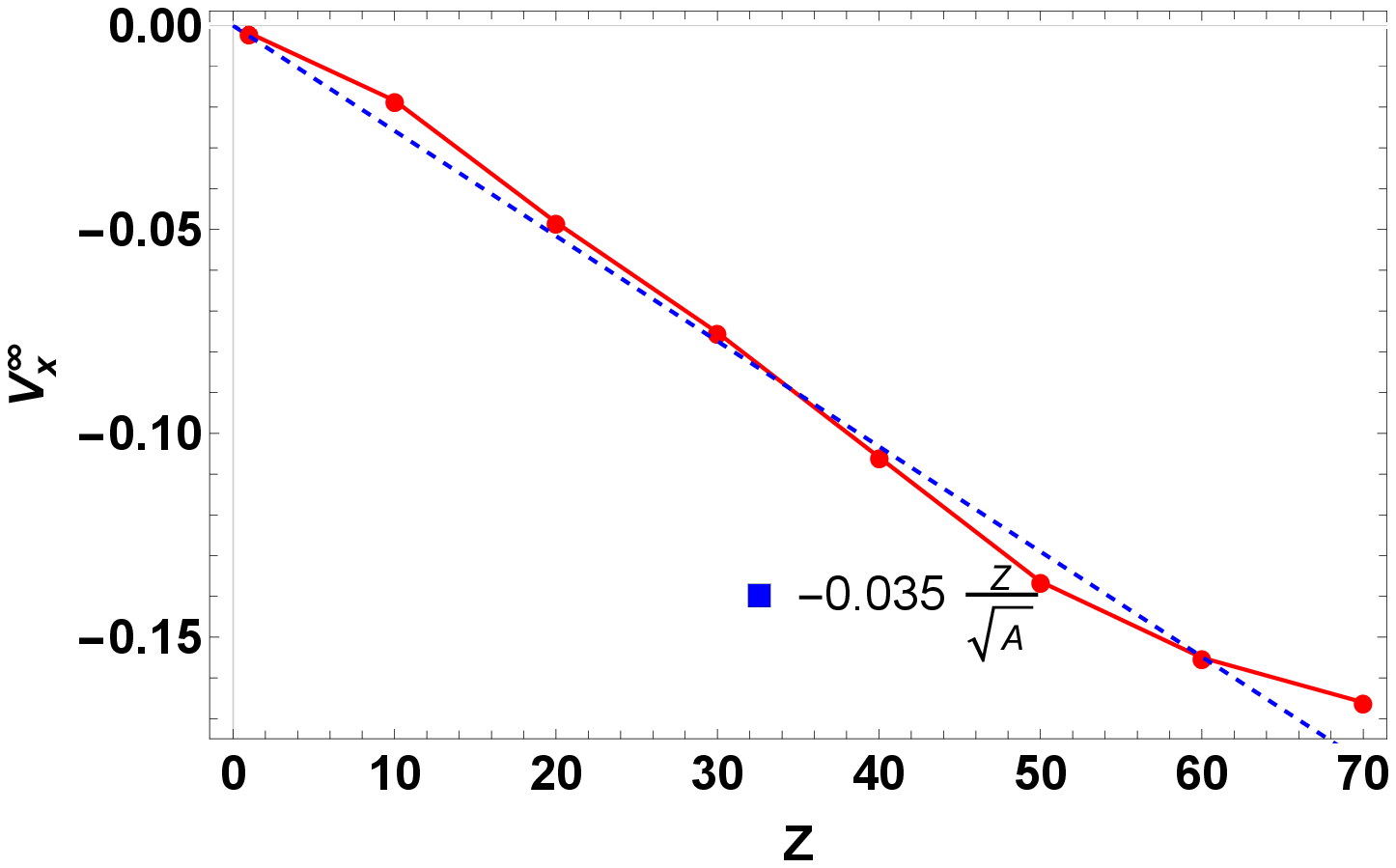} 
\includegraphics[height=5.2cm]{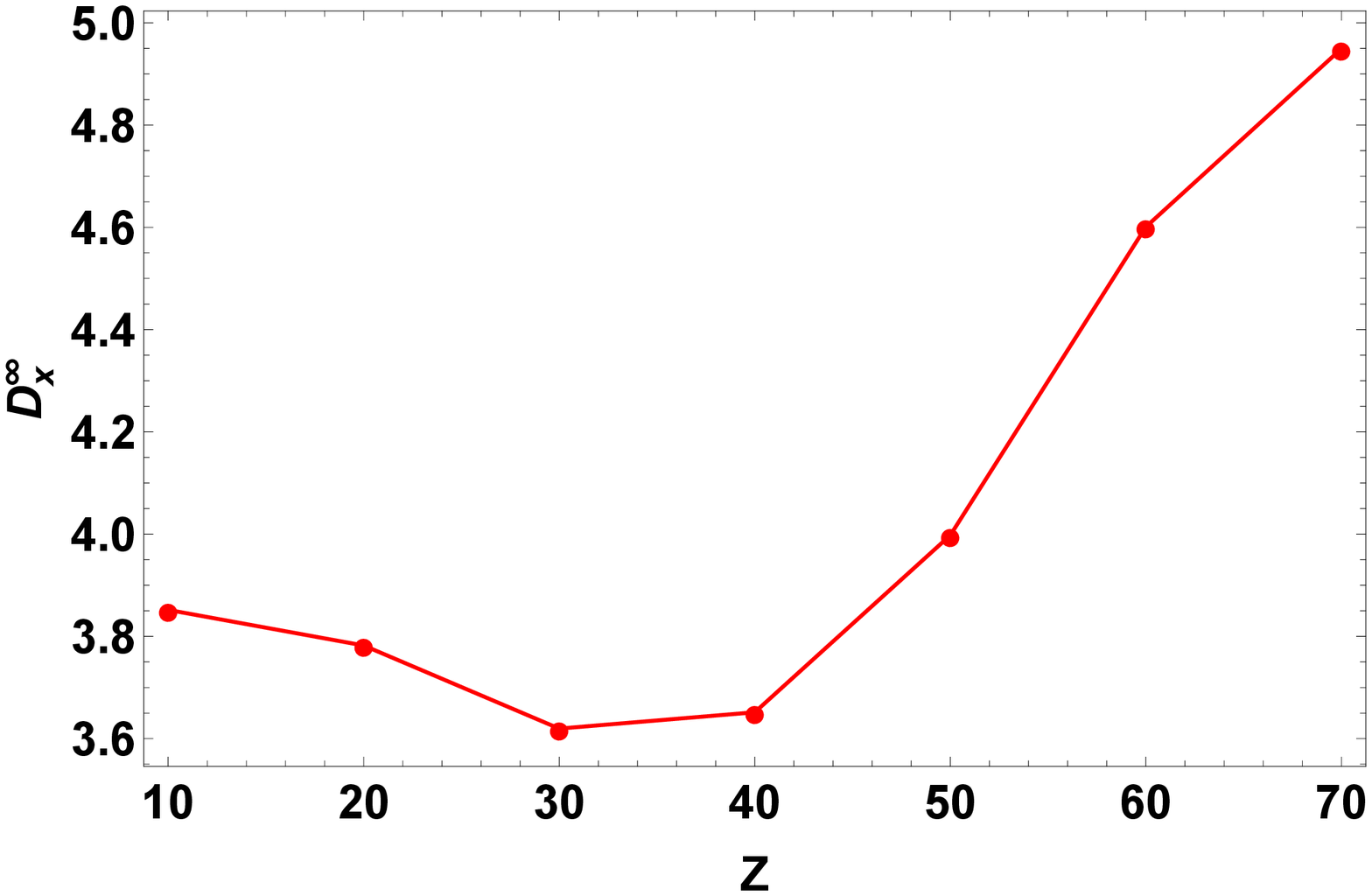}}
\caption{W transport dependence on the ionization rate $Z$ : the asymptotic
pinch velocity (left panel) and the diffusion coefficient (right panel). }
\label{Figure 5}
\end{figure}

The mechanism of generation of the radial pinch depends essentially on the
product $Z\Phi .$ Thus, the ionization rate has a similar effect with the
amplitude $\Phi $ of the turbulence. As seen in Fig. \ref{Figure 5} (left
panel), the pinch velocity has an approximately linear increase followed by
the tendency of saturation, a behaviour that is similar to the dependence on 
$\Phi $ (Fig. \ref{Figure 4} (left panel)). The diffusion coefficient shown
in \ref{Figure 5} (right panel) has a more complicated dependence on $Z,$
but the variation of $D_{x}^{\infty }$\ on the relevant range of $Z$\ is
small (of the order $\pm 20\%$ of the average). The influence of $Z$ on the
diffusion is produced through the effective parallel decorrelation time $%
\tau _{z}^{\infty }$ that depends on $Z$.\ \ 

\begin{figure}[tbh]
\centerline{\includegraphics[height=5.2cm]{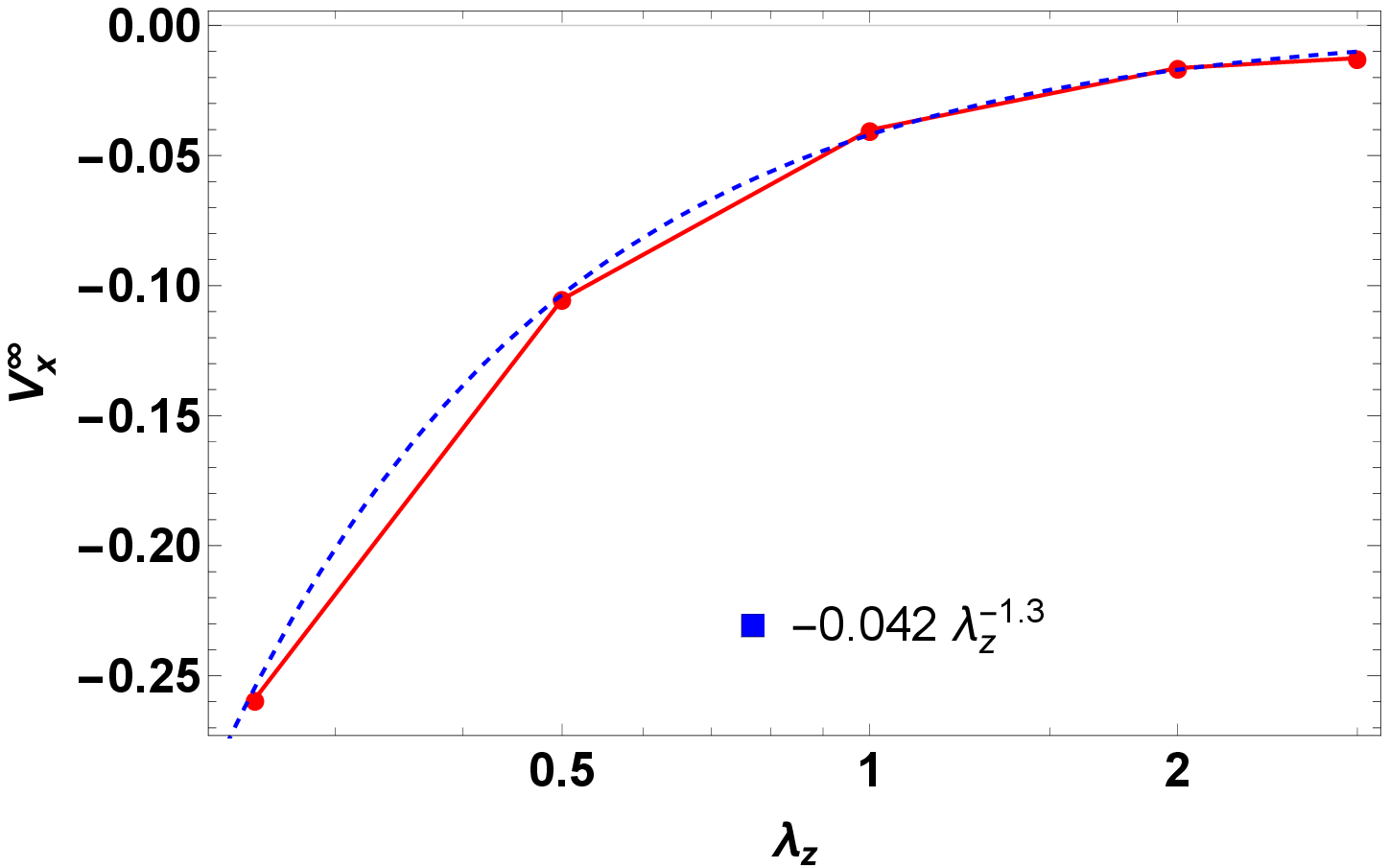} 
\includegraphics[height=5.2cm]{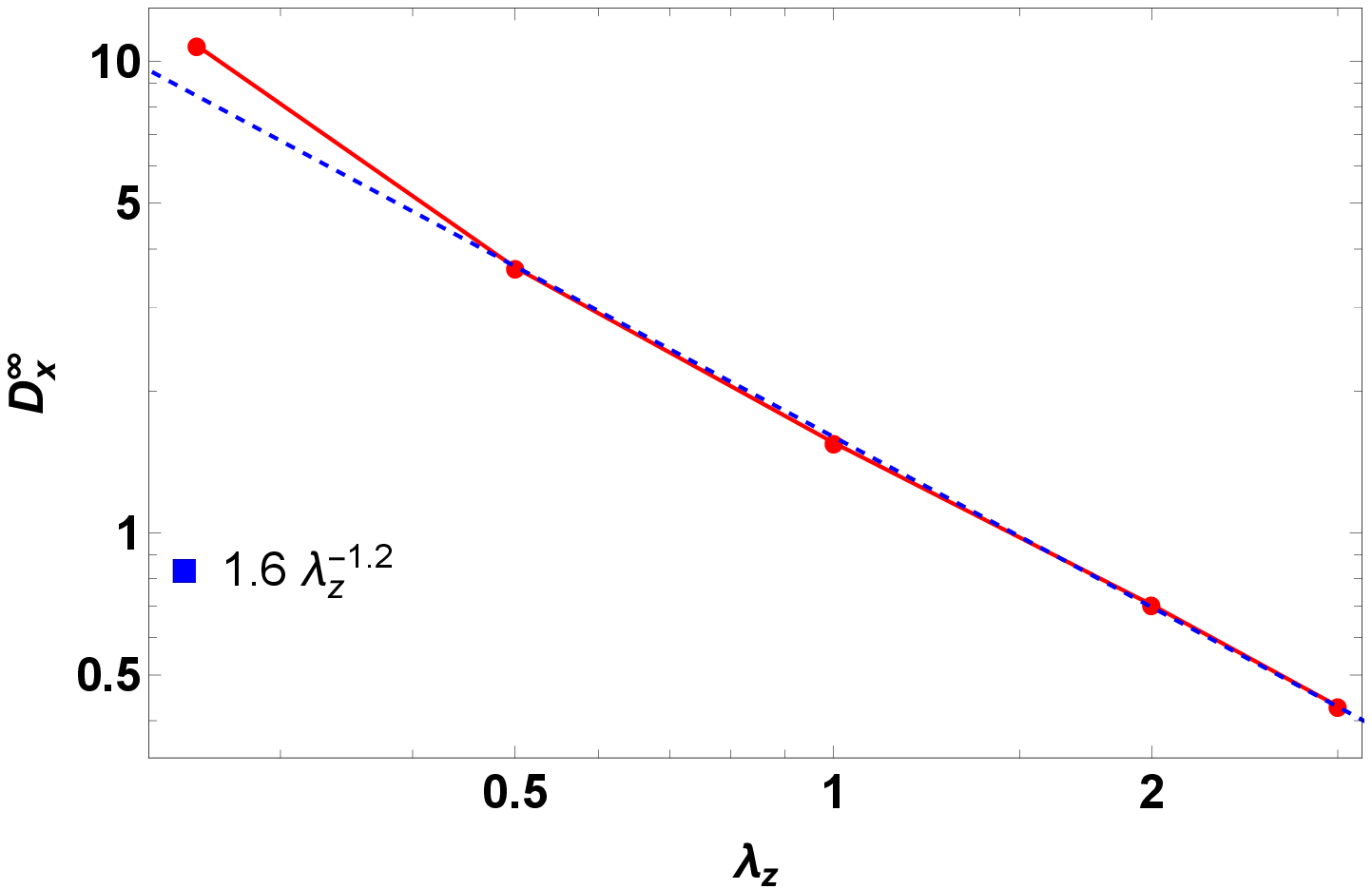}}
\caption{W transport dependence on the parallel correlation length $\protect%
\lambda_{z}$ : the asymptotic pinch velocity (left panel) and the diffusion
coefficient (right panel). }
\label{Figure 6}
\end{figure}
\bigskip

\begin{itemize}
\item Parallel correlation length $\lambda _{z}$
\end{itemize}

The pinch mechanism analyzed here appears only in three-dimensional
stochastic potentials. But, as discussed in Section 5, $V_{x}^{\infty }$
essentially results from the symmetry breaking of the HDs determined by the
energy conservation. The potential energy does not dependent on $\lambda
_{z} $, which means that a finite $\lambda _{z}$\ is necessary, but its
direct quantitative influence on the pinch mechanism is small.

However, $\lambda _{z}$ has a strong influence on the transport through the
parallel decorrelation time $\tau _{z}^{\infty }$ in Eq. (\ref{tauz}) that
increases with $\lambda _{z}$ faster than linearly. It explains the large
decrease rate of both $V_{x}^{\infty }$ and $D_{x}^{\infty }$ seen in Fig. %
\ref{Figure 5}, which shows that $\left\vert V_{x}^{\infty }\right\vert \sim
\lambda _{z}^{-1.3}$ and $D_{x}^{\infty }\sim \lambda _{z}^{-1.2}.$

Thus, $\lambda _{z}$ determines the decrease of $V_{x}^{\infty }$ and $%
D_{x}^{\infty }$ only through the modification of the $\tau _{z}^{\infty }$%
.\ \ 

\begin{itemize}
\item Poloidal velocity $V_{p}$
\end{itemize}

\begin{figure}[tbh]
\centerline{\includegraphics[height=5.2cm]{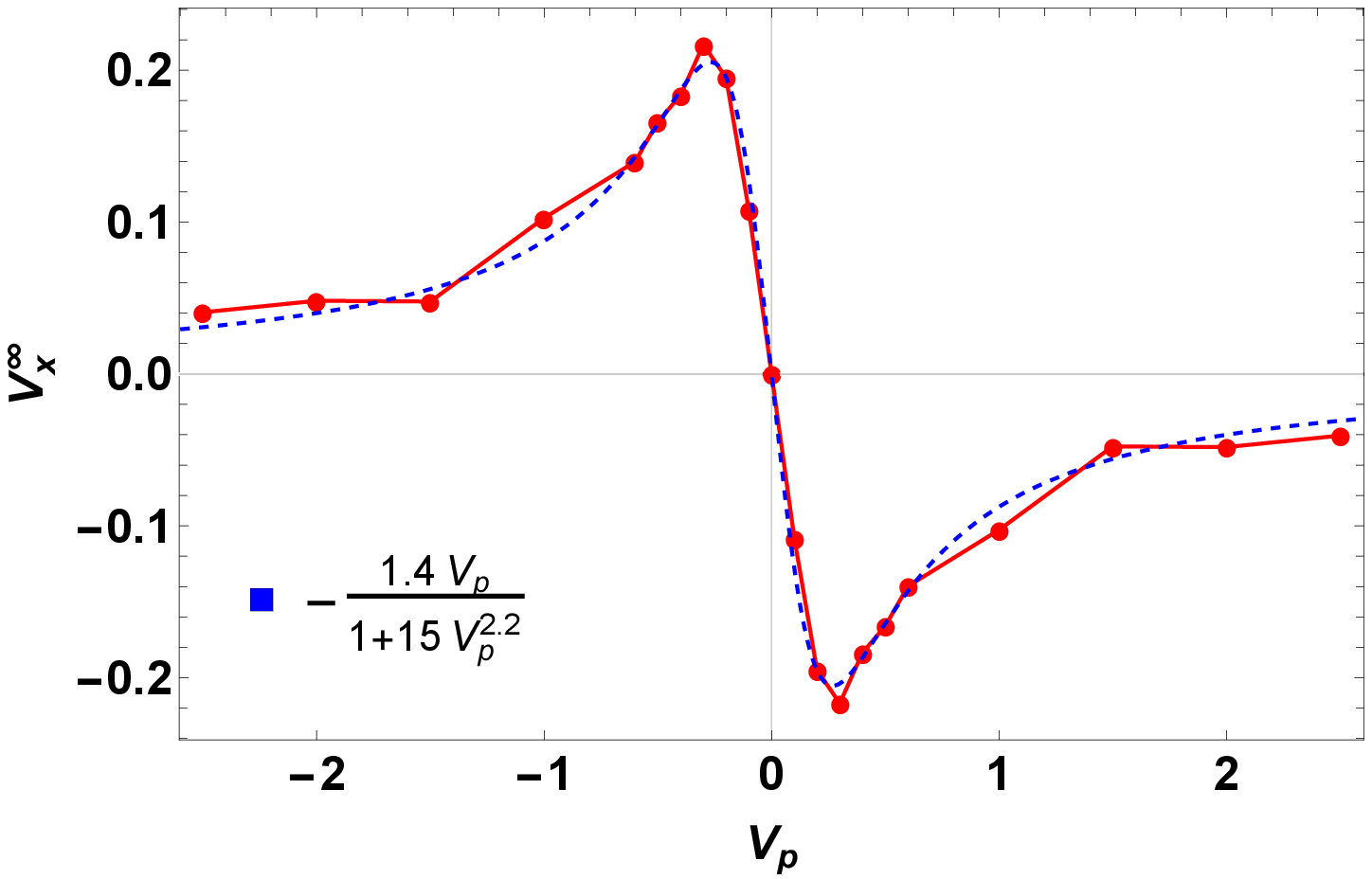} 
\includegraphics[height=5.2cm]{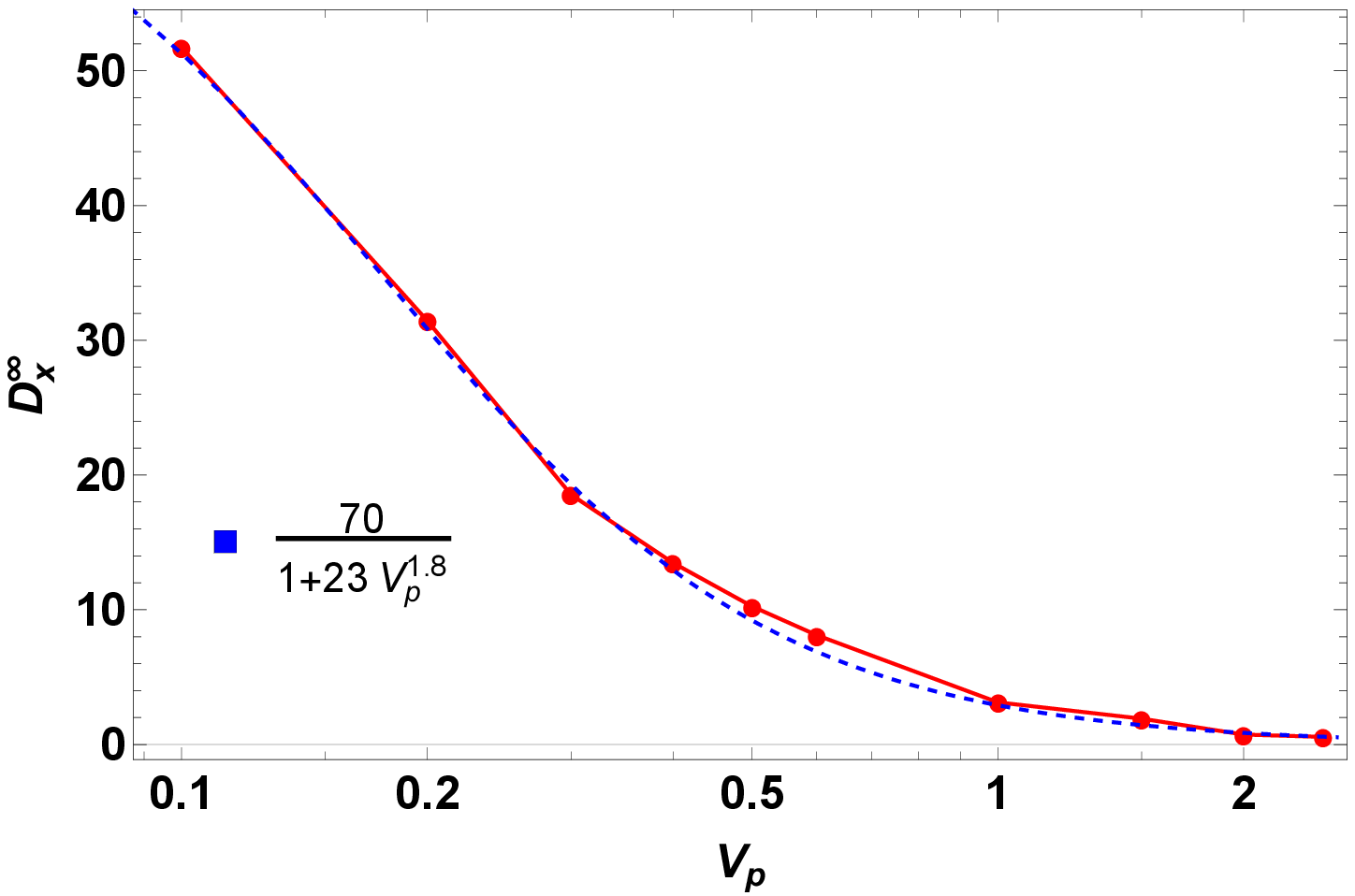}}
\caption{W transport dependence on the poloidal velocity $V_{p}$ : the
asymptotic pinch velocity (left panel) and the diffusion coefficient (right
panel). }
\label{Figure 7}
\end{figure}

The poloidal average velocity is the source of the hidden drifts. It has a
strong influence on both the pinch velocity and the diffusion coefficient,
as seen in Fig. \ref{Figure 7}.

The equations of motion (\ref{eqm})-(\ref{eqmz}) are invariant at\ the
change $V_{p}\rightarrow -V_{p}$ and $\mathbf{x}\rightarrow -\mathbf{x}$,
which implies that the conditional displacements and the HDs change their
sign when $V_{p}\rightarrow -V_{p}.$ Thus, the pinch velocity $V_{x}^{\infty
}$\ is an anti-symmetrical function of $V_{p},$ as seen in Fig. \ref{Figure
7} (left panel). $V_{x}^{\infty }$\ is linear in $V_{p}$ at small $V_{p},$
it has a maximum at $V_{p}\cong 0.3$ and a long tail with $V_{x}^{\infty
}\sim V_{p}^{-1.2}$ at large $V_{p}.$

The direction of the pinch produced by the parallel acceleration can be
changed from inward to outward by inversing the orientation of the poloidal
velocity.

The diffusion coefficient is strongly influenced by $V_{p},$ which
determines a large decrease of $D_{x}^{\infty },$ as seen in Fig. \ref%
{Figure 7} (right panel).

Thus, $V_{p}$ has a special effect on the transport, different compared to
the other parameters. The pinch velocity $V_{x}^{\infty }$\ is modified
because $V_{p}$ influences the amplitude of the HDs. The diffusion
coefficient $D_{x}^{\infty }$ is modified because $V_{p}$ influences the
structure of the contour lines of the total potential.\ \ \ \ \ \ 

\begin{itemize}
\item Time decorrelation $\tau _{d}$
\end{itemize}

Our results have confirmed the idea that the time variation of the
stochastic potential determines a process of elimination of the pinch
velocity by strengthening the random aspects of the motion. In addition to
this, the Lagrangian energy is not a constant, but a fluctuating function of
time.

\begin{figure}[tbh]
\centerline{\includegraphics[height=5.2cm]{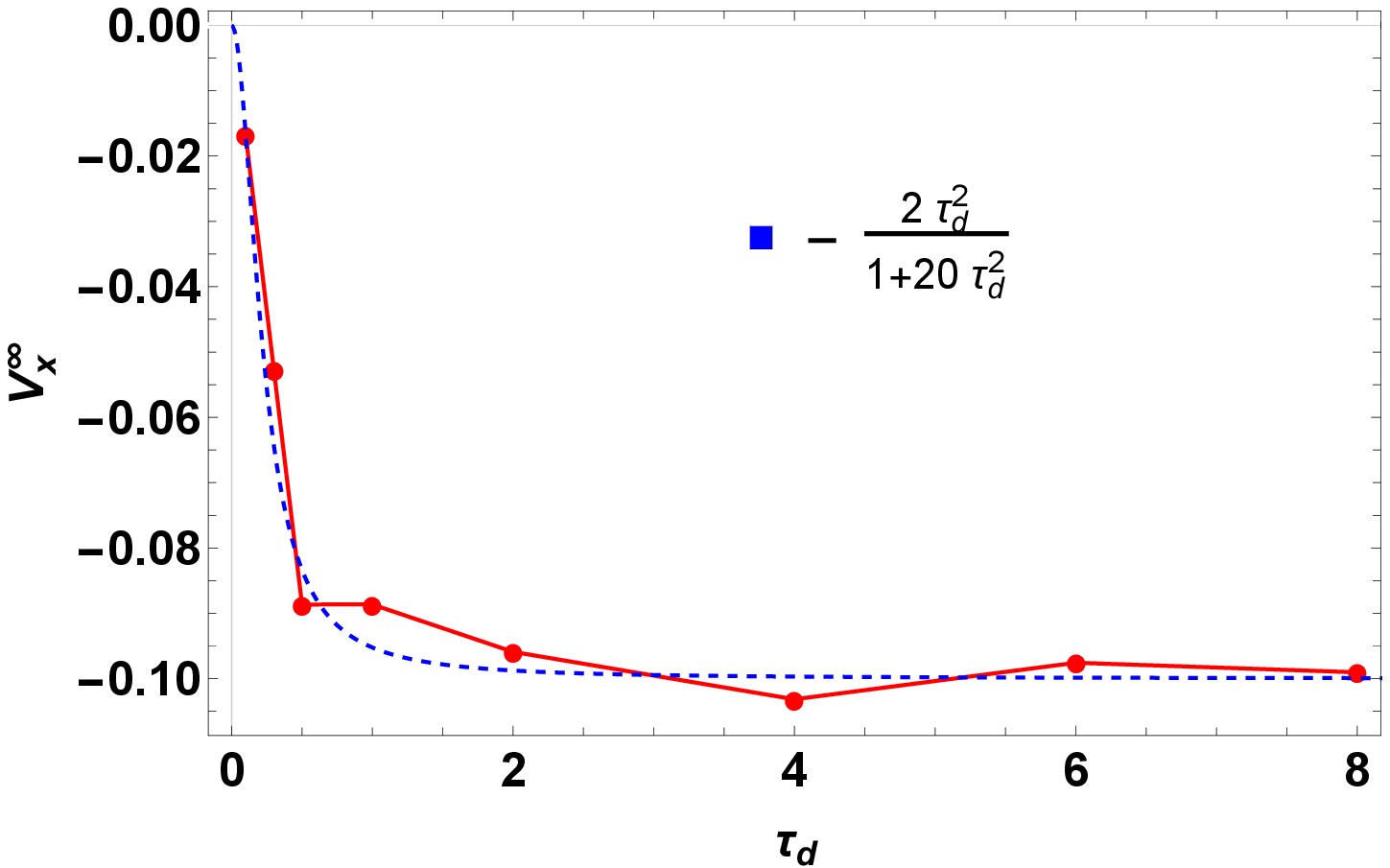} 
\includegraphics[height=5.2cm]{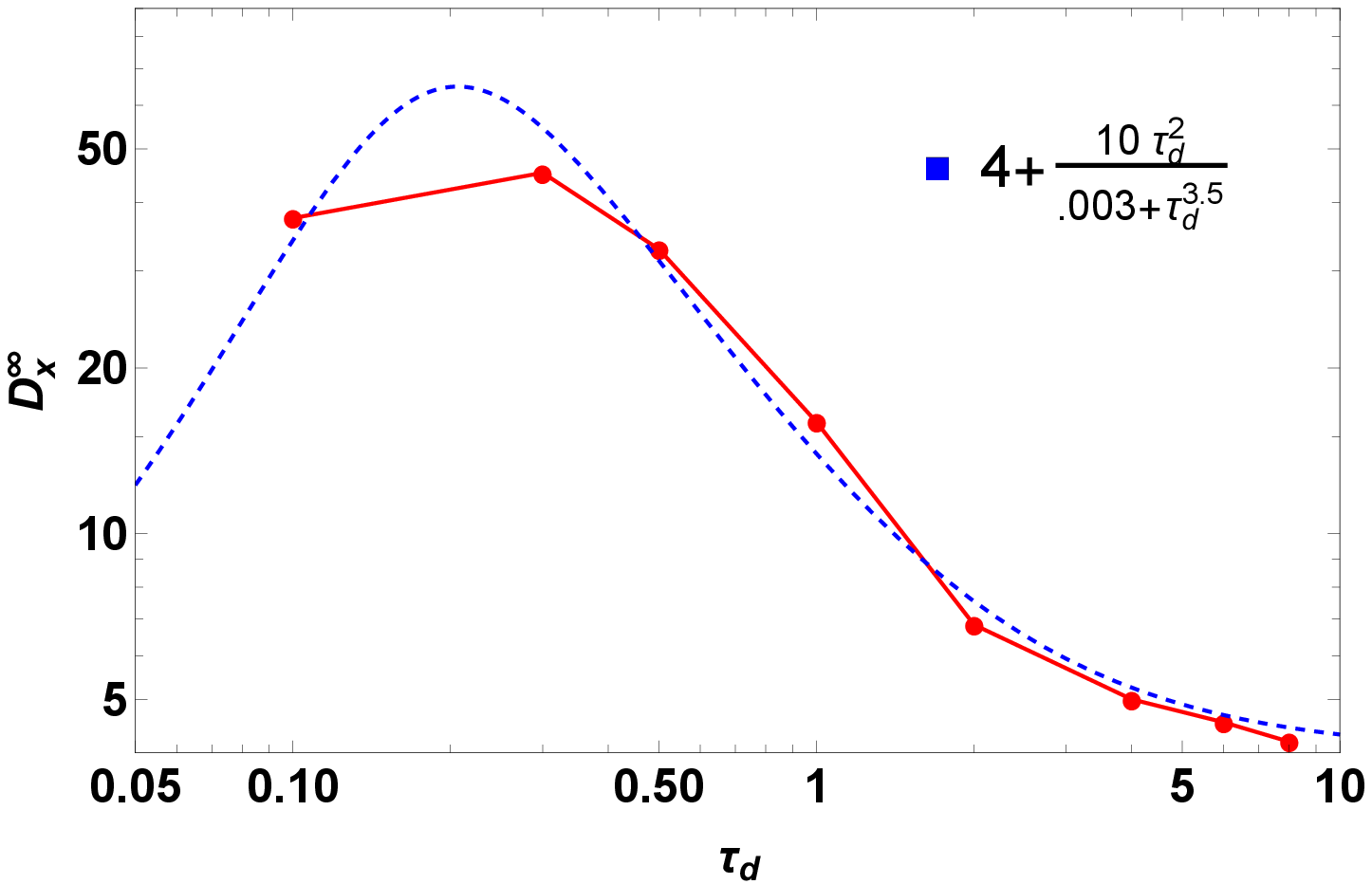}}
\caption{W transport dependence on the time variation parameter $\protect%
\tau _{d}$ : the asymptotic pinch velocity (left panel) and the diffusion
coefficient (right panel). }
\label{Figure 10}
\end{figure}

However, the pinch velocity survives in time dependent potentials $\phi (%
\mathbf{x},z,t)$ if the\ decorrelation tine $\tau _{d}$ is not too small. As
seen in Fig. \ref{Figure 10} (left panel), $V_{x}^{\infty }$ is weakly
dependent on $\tau _{d}$ for $\tau _{d}>1,$\ and it has fast decrease as $%
V_{x}^{\infty }\sim -\tau _{d}^{2}$ for $\tau _{d}<0.5.$ The pinch velocity
is eliminated for fast time variation with $\tau _{d}\ll \tau _{fl}.$

A different behaviour was obtained for $D_{x}^{\infty }.$ As seen in Fig. %
\ref{Figure 10} (right panel), $D_{x}^{\infty }$ increases at small $\tau
_{d}$\ (in the quasi-linear regime, $\tau _{d}\ll \tau _{fl}$), reaches a
maximum and decreases due to trapping for $\tau _{d}\gg \tau _{fl}.$\ At
larger $\tau _{d},$ of the order of the parallel decorrelation time,\ $%
D_{x}^{\infty }$\ saturates at the value corresponding to the static
potential. This behaviour results from the combination of the time
decorrelation processes produced by the time variation of the potential
(represented by the time dependence of the EC (\ref{exp-dec})) and by the
parallel motion (\ref{tauef}). This yields an effective decorrelation time $%
\tau ^{\infty }$ that is $\tau ^{\infty }\cong \tau _{d}$ for $\tau _{d}\ll
\tau _{z}^{\infty }$\ and $\tau ^{\infty }\cong \tau _{z}^{\infty }$ for $%
\tau _{d}\gtrsim \tau _{z}^{\infty }.$\ 

\begin{itemize}
\item Energy of the W ions
\end{itemize}

The energy is directly connected to the mechanism of pinch generation. The
cut of the conditional average displacements $\left\langle \mathbf{x}%
(t)\right\rangle _{\phi ^{0}},$ which has the dominant influence of the
asymmetry of the HDs, appears at $\phi _{\max }=W/(Z\Phi ).$ It is expected
that the change of $W$ determines a variation of the pinch velocity of the
order of those produced by $Z$ or $\Phi ,$\ but inverse, in the sense that $%
\left\vert V_{x}^{\infty }\right\vert $ is a decreasing function of $W.$\ 

We have obtained only a very weak decrease of the pinch velocity with the
energy.

The reason is the dependence of the parallel decorrelation time $\tau
_{z}^{\infty },$ which is a decreasing function of $W.$ The saturation of $%
V_{x}(t)$ at a smaller time determines the increase of $\left\vert
V_{x}^{\infty }\right\vert $ (because it is a decreasing function of $t).$
The two effects of the parallel acceleration (symmetry breaking of the HDs
and decorrelation) are opposite in this case of the energy dependence, and
they partly compensate.

The increase of the energy also determines a weak increase of the diffusion
coefficient. \ 

\section{Discussions and conclusions}

The main finding of this work is a radial pinch that is generated by the
stochastic parallel acceleration in turbulent plasmas. It is significant for
high $Z$ impurities and negligible for plasma ions. We have shown that the
pinch is produced in three-dimensional turbulence by the interaction of the
parallel motion with the HDs, a special type of quasi-coherent radial motion
that appears due to a poloidal average velocity.

We have also shown that the influence of the parallel motion on the
transport through the parallel decorrelation time $\tau _{z}^{\infty }$ is
much stronger for heavy impurities than for plasma ions. The fluctuations of
the parallel velocity are very large for W ions, and they determine a
smaller parallel decorrelation time that depends on the parameters of the
parallel motion $\tau _{z}^{\infty }(W,Z,\Phi ,\lambda _{z}).$ This complex
decorrelation process influences both the pinch velocity and the diffusion
coefficient. It leads to an unusual diffusion regime of super-Bohm type and
modifies the scaling laws of $V_{x}^{\infty }$\ and $D_{x}^{\infty }.$\ 

The physical domains of the main parameters of the transport model were
explored for evaluating the scaling laws and for obtaining the range of the
normalized pinch velocity and diffusion coefficient. The typical values of $%
\left\vert V_{x}^{\infty }\right\vert $\ are in the interval $(0.05,~0.25).$%
\ \ \ \ \ \ 

We underline that the dependence of $V_{x}^{\infty }$ on $V_{p}$\ (Fig. \ref%
{Figure 7} (left panel)) provides a very efficient control possibility. The
change of $\mathbf{V}_{p}$ from the direction of the electron to the ion
diamagnetic velocity determines the inversion of the pinch from inward to
outward direction. A strong variation of $V_{x}^{\infty }$\ with $V_{p}$\
exists at small $\left\vert V_{p}\right\vert \lesssim 0.5,$\ which shows a
high sensitivity of the pinch velocity to the poloidal velocity.\ \ \ \ 

The relevance of the pinch velocity for ASDEX Upgrade, JET and ITER can be
evaluated from the dimensionless results. The main difference (concerning $%
V_{x}^{\infty })$ between present plasmas and ITER is the electron
temperature. Due to the time-scale separation of the atomic and transport
processes, the W impurities are in coronal equilibrium. The fractional
abundance of each ionization stage is a function of the electron temperature
that is practically not influenced by the transport \cite{Putterich2008}.
This determines different ranges of the ionization rates for the present
plasmas and ITER. In the first case $Z$ varies from boundary to the center
in the interval $(20,~48),$ while in the second case the interval is $%
(45,~63).$ Typical values of $Z$ in the core plasma are $Z=[43,~44,~57],$\
where the first value in this and the following triads corresponds to ASDEX
Upgrade, the second to JET and the third to ITER. This determines normalized
values of the pinch velocity of the order $\left\vert V_{x}^{\infty
}\right\vert \approx \lbrack 0.10,~0.11,~0.15].$\ Using typical parameters
of these plasmas, the pinch velocities are of the order $\left\vert
V_{x}^{\infty }\right\vert \approx \lbrack 160,~110,~194]~m/sec.$ Thus, the
pinch velocity is larger in the ITER plasmas roughly by $50\%$\ at similar
parameters of the turbulence and poloidal velocity. The convection time to
plasma center is very small $\Delta t_{c}=a/V_{x}^{\infty }\approx \lbrack
4,~11,~10]~msec.$\ Convection dominates diffusion in all cases, because $%
\Delta t_{c}\ll \Delta t_{dif},$\ where $\Delta
t_{dif}=a^{2}/(2D_{x}^{\infty })$ is the diffusive time. The ratio $r=\Delta
t_{c}/\Delta t_{dif}$ is $r\approx \lbrack 0.18,~0.10,~0.07].$

The above estimation shows a very strong effect of the acceleration induced
pinch on the dynamics of the W impurities. This enables the idea that the
processes found here remain significant in a frame of a realistic transport
model that includes W ion collisions, the polarization drift and
neoclassical aspects. The future work will be dedicated to the development
of the model and to the examination of the of interaction of $V_{x}^{\infty
} $ with other types of radial pinches.

In conclusion, this study provides understanding of the complex processes of
interaction of the parallel acceleration with the perpendicular transport.
The main effect consists of the generation of a radial pinch that appears to
be significant for the dynamics of the W impurities in present plasmas and
ITER.

\bigskip

\textit{Acknowledgements} This work has been carried out within the
framework of the EUROfusion Consortium and has received funding from the
Euratom research and training programme 2014-2018 and 2019-2020 under grant
agreement No 633053 and from the Romanian Ministry of Research and
Innovation. The views and opinions expressed herein do not necessarily
reflect those of the European Commission.

\end{document}